\documentclass[12]{article}
\usepackage[utf8]{inputenc}
\usepackage{amsmath, amsfonts}
\usepackage[margin=1.0in]{geometry}
\usepackage{algorithm}
\usepackage[noend]{algpseudocode}
\usepackage{graphics, graphicx}
\usepackage{authblk}
\usepackage{xcolor}
\usepackage{hyperref}

\usepackage{natbib}

\DeclareMathOperator{\EX}{\mathbb{E}}

\usepackage{tabu}

\title{Latent Network Estimation and Variable Selection for Compositional Data via Variational EM}

\author[1]{Nathan Osborne}
\author[2]{Christine B. Peterson}
\author[1]{Marina Vannucci}
\affil[1]{Department of Statistics, Rice University, Houston, TX}
\affil[2]{Department of Biostatistics, The University of Texas MD Anderson Cancer Center, Houston, TX}

\begin{document}

\maketitle

\vspace{-20pt}
\begin{abstract}
Network estimation and variable selection have been extensively studied in the statistical literature, but only recently have those two challenges been addressed simultaneously. In this paper, we seek to develop a novel method to simultaneously estimate network interactions and associations to relevant covariates for count data, and specifically for compositional data, which have a fixed sum constraint. We use a hierarchical Bayesian model with latent layers and employ spike-and-slab priors for both edge and covariate selection. For posterior inference, we develop a novel variational inference scheme with an expectation maximization step, to enable efficient estimation. Through simulation studies, we demonstrate that the proposed model outperforms existing methods in its accuracy of network recovery. We show the practical utility of our model via an application to microbiome data. The human microbiome has been shown to contribute to many of the functions of the human body, and also to be linked with a number of diseases. In our application, we seek to better understand the interaction between microbes and relevant covariates, as well as the interaction of microbes with each other. We call our algorithm SINC (Simultaneous Inference for Networks and Covariates) and provide a Python implementation, which is available online.

\noindent    
\textbf{Keywords:} graphical model, variational inference, EM algorithm, count data, Bayesian hierarchical model, microbiome data
  
\end{abstract}

\vspace{-10pt}
\section{Introduction}
Variable selection, also known as feature selection, is a well-studied subject in the statistical literature, particularly in the context of regression models, where many approaches have been proposed. Feature selection offers an opportunity to both improve model predictions, by avoiding the inclusion of noisy or irrelevant predictors, and to identify interpretable parsimonious models. Frequentist approaches often use a penalized likelihood to obtain sparse estimates of the regression coefficients, and include methods such as LASSO \citep{LASSO}, adaptive LASSO \citep{Adaptive}, and SCAD \citep{SCAD}. Alternatively, Bayesian approaches employ carefully constructed priors on the regression coefficients to identify the relevant variables. Spike-and-slab priors, first proposed by \cite{mitchell_beauchamp_1988}, are a popular class of priors that use a latent indicator to represent variable inclusion. Conditional on the indicators, the regression coefficients are assumed to come from a mixture prior representing important vs negligible effects  \citep{GeorgeMcCulluch, brown98}. In addition to sparse estimation of the coefficients, these priors produce posterior probabilities of inclusion (PPIs) for each covariate that capture the uncertainty in the selection. Spike-and-slab priors have been extended to regression models for non-Gaussian data, including binary, multinomial and count responses \citep{Raftery1996,Ntzoufras2003, Sha2004, Wadsworth, MicroBVS}.

A parallel development has happened in the  graphical model literature: in this framework, nodes correspond to variables, and edges connecting these nodes represent conditional dependence relations. In the Gaussian setting, the problem of selecting edges in the graph reduces to the estimation of a sparse inverse covariance matrix, since exact zeros in this matrix, which is also known as the precision matrix, correspond to conditional independence relations \citep{Dempster}. In frequentist settings, penalized likelihood methods, such as neighborhood selection \citep{Meinshausen} and the graphical LASSO \citep{Yuan2007,Friedman2008}, have been proposed. These methods have been extended to count data by using data transformations \citep{spiecEasi} or penalized log-likelihood methods \citep{gCoda}. In Bayesian inference, the G-Wishart prior \citep{Roverato}, which is the conjugate prior that imposes exact zeros in the precision matrix, has been explored by several authors for inference in Gaussian graphical models, but poses significant computational challenges \citep{Lenkoski2011}. As a result, this prior is not easily scalable. Alternative shrinkage constructions that employ continuous priors on the off-diagonal elements of the precision matrix have been proposed, such as the Bayesian graphical lasso \citep{WangBGLasso}, which relies on double exponential priors, and mixture priors \citep{Wang2015}, inspired by the spike-and-slab priors in the regression framework discussed above. To enable estimation of the spike-and-slab model of \cite{Wang2015} in high-dimensional settings, \cite{Li2017} recently proposed an efficient expectation conditional maximization method, which offers an attractive alternative to stochastic search approaches.

In this paper, we propose a novel Bayesian hierarchical model for count data that allows for simultaneous estimation of covariate dependence and network interactions. Methods for simultaneous estimation are gaining popularity, with approaches including penalized likelihood methods \citep{Rothman2010,mLDM}, and, most recently, spike-and-slab lasso prior models \citep{Deshpande2019}. By accounting for covariate selection, simultaneous estimation methods are able to control for those variables, which ultimately leads to more accurate network estimation. Moreover, simultaneous estimation can improve the detection of covariate effects, as noted by  \cite{Deshpande2019}.  However, with the exception of \cite{mLDM}, these methods are not suitable for count data.  In our approach, we consider multivariate count data, and specifically compositional data that have a fixed sum constraint. We model the data using a Dirichlet-Multinomial likelihood and then introduce a latent layer by modeling the log concentration parameters via a Gaussian distribution. We account for covariates through the mean function of the latent layer and employ multivariate variable selection spike-and-slab priors that allow each covariate to be relevant for individual response variables \citep{richardson2010, stingo2010}. We also capture a network of latent dependence relationships by estimating the inverse covariance matrix via the mixture prior of \cite{Wang2015}. For posterior inference, we implement a novel variational Bayes approach that includes an expectation-minimization (EM) step to estimate the model. This allows us to gain flexibility by using a Bayesian model, while still remaining computationally efficient. Additionally, the algorithm is developed so that multiple steps can be run in parallel, achieving larger computational gains. We show through simulations that our method outperforms the LASSO-based approach of \cite{mLDM}. We refer to our model as SINC (Simultaneous Inference of Networks and Covariates).

Compositional data are often collected in chemistry, geology, and biology applications.  In biomedicine, modern genomic sequencing technologies have allowed investigators to collect samples on the human microbiome. Microbes associated with the human body include eukaryotes, archaea, bacteria, and viruses, which have been shown to contribute to important bodily functions including food digestion and energy supply.
The human microbiome has also been implicated in many diseases including colorectal cancer, inflammatory bowel disease, and immunologically mediated skin diseases.  The observed data from a microbiome study are typically short reads of DNA sequences, which are clustered to create operational taxonomic units (OTUs). The abundances across samples of these OTUs, which represent genetically close groups of microbes assumed to have similar functions, are taken as input to downstream analysis. A challenge to modeling these data is that the number of counts for a particular OTU depends on the number of sequences collected for that sample, meaning that the observed counts are dependent on each other, as they constitute proportions of a whole. This results in data that are compositional. For these reasons, Dirichlet-Multinomial distributions are particularly appropriate to model microbiome data, as demonstrated by several authors \citep{chen_li_2013, LiMa,Wadsworth}.  In the application of this paper we focus on two questions of interest in the understanding of the microbiome: which variables influence the microbial abundances, and what are the dependence relationships among microbes.  The abundance of microbes or groups of microbes is dependent on many factors.  Microbial abundance may be related to  external covariates, such as diet, cytokines, or use of medication.These factors influence the microbiome by introducing new organisms, changing the abundance of metabolites, or altering the pH of their environment. For example, consumption of an animal-based diet high in meat has been shown to increase production of bile acid, which inhibits growth of bacteria belonging to the Bacteroidetes and Firmicutes phyla \citep{David2014}. Antibiotics can alter the microbiome substantially, by killing off components of the microbiome in addition to the bacteria triggering the infection \citep{Edwards2019}. As we understand more about the importance of the microbiome, it is also critical to understand what factors lead to the prevalence of different microbes.  Here we apply the proposed method to real data from the Multi-\textsf{'}Omic Microbiome Study-Pregnancy Initiative (MOMS-PI) study, to estimate the interaction between microbes in the vagina, as well as the interplay between vaginal cytokines and microbial abundances, providing insight into mechanisms of host-microbial interaction during pregnancy.

The paper is outlined as follows: in Section \ref{section:Model}, we describe the proposed hierarchical model, followed by the variational EM estimation method in Section \ref{section:comp}. We  provide a simulation study in section \ref{section:sim}, and then showcase the proposed model in an application to multi-omic data from a study on the role of the microbiome in pregnancy in Section \ref{section:app}. Finally, we discuss the advantages of the proposed model in Section \ref{section:disc}.

\section{Proposed Model}
\label{section:Model}

Suppose we have observed multivariate counts arranged in an $n \times p$ matrix, $\mathbf{X}$, where $p$ is the number of observed variables measured across $n$ samples. We then let the $p$-vector $\mathbf{X}_i$ correspond to the measurements for observation $i$, and the matrix entry $x_{i,j}$ correspond to the $j^{th}$ variable measurement for the $i^{th}$ observation. We also observe $q$ covariates for each of the $n$ observations, with these $q$ additional factors possibly influencing the measured counts for each observation. We arrange the covariate data in an $n$ $\times$ $q$ matrix, $\mathbf{M}$.

We are interested in understanding the conditional dependence relationships among the $p$ variables while simultaneously selecting the relevant covariates. We adopt a hierarchical model formulation with a latent Gaussian layer, similarly to \cite{mLDM}, as 
 \begin{align}
 \begin{split}
    \mathbf{Z}_i \mid \mathbf{B}_0, \mathbf{M}_i, \mathbf{B}, \mathbf{\Omega} & \sim \text{MVNorm}(\mathbf{B}_0 + \mathbf{M}_i \mathbf{B}, \mathbf{\Omega}^{-1}) \\
    \boldsymbol{\alpha}_i & = \exp \{ \mathbf{Z}_i \} \\
    \mathbf{h}_i \mid \boldsymbol{\alpha}_i & \sim \text{Dirichlet}(\boldsymbol{\alpha}_i) \\
    \mathbf{X}_i \mid \mathbf{h}_i & \sim \text{Multinomial}(\mathbf{h}_i, {n_i}). 
 \end{split}
 \label{eq:model}
\end{align} 

\noindent In this hierarchical formulation, we introduce a latent normal variable $\mathbf{Z}_i$, which is a direct transformation of the concentration parameter $\boldsymbol{\alpha}_i$ and therefore controls the observed counts $\mathbf{X}_i$. This model has several important features: the Dirichlet-Multinomial likelihood for count data, $\mathbf{X}_i$, allows us to account for overdispersion as well as the compositional nature of the data. The dependence on covariates is incorporated through the mean of the multivariate normal, where the observed covariates $\mathbf{M}_i$ have effects $\mathbf{B}$. The dependence among the $\mathbf{Z}_i$ is captured by the inverse covariance matrix, also known as the precision matrix, $\mathbf{\Omega}$. The 1 $\times p$ vector $\mathbf{B}_0$ accounts for the mean of each column of the latent matrix $\mathbf{Z}$. In our modeling approach, careful consideration of the priors on the covariate effects $\mathbf{B}$, the intercepts $\mathbf{B}_0$ and the precision matrix $\mathbf{\Omega}$ allows us to construct a directed graph between covariates $\mathbf{M}$ and latent variables $\mathbf{Z}$, as well as an undirected graph between the columns of $\mathbf{Z}$. 

For microbiome studies, \cite{Gloor} noted that the observed compositional data have a different correlation structure than the true underlying abundances. More specifically, due to the fixed sum constraint, compositional data tend to exhibit negative correlations.  In model formulation \eqref{eq:model}, we interpret the latent layer $\textbf{h}_{i}$ to be the relative abundances, and $\boldsymbol{\alpha}_{i}$ to be the absolute abundances \citep{mLDM}. By estimating a network on the latent \textbf{Z}, we capture the network of the underlying, absolute abundances through the precision matrix $\mathbf{\Omega}$. Therefore, even though the latent Gaussian layer does not allow us to recover relationships directly among the observed counts, the inferred dependences do  provide some insights into the relationships among the underlying processes.  Latent graphical models for Poisson-distributed count data that use Gaussian layers were used by \cite{Vinci18}, for spike-count data. See also \cite{Talhouk12} and \cite{LiMcCormick20} for latent graphical model constructions for binary data.

\subsection{Prior on covariate effects \textbf{B}}

Here we describe the prior on the covariate effects, which enables selection of the important associations between $\mathbf{X}$ and other potentially related factors $\mathbf{M}$. We consider the effects of the covariates $\textbf{M}$ on each column of $\textbf{Z}$ separately, which means that we will be able to update the columns of $\textbf{B}$ independently of each other. Here $\mathbf{B}$ is a $q \times p$ matrix, where each column of $\textbf{B}$ represents the vector of regression coefficients for the $q$ covariates of $\textbf{M}$ on the $j^{th}$ column of $\textbf{Z}$.  We use a spike-and-slab prior on each element of the matrix $\mathbf{B}$, which shrinks features that do not influence $\mathbf{Z}$ to zero. Remember that we are looking at the columns of $\textbf{Z}$ one at a time, and can thus say that any entry from the $j^{th}$ column, $Z_{i,j}$, comes from a $\text{Normal}(\mathbf{B}_{0j} + \mathbf{M}_i \mathbf{B}_j,\sigma_j^*)$, where $\sigma_j^*$ is the standard deviation of the $j^{th}$ column of $\mathbf{Z}$, found by using the properties of the multivariate normal distribution shown in equation (\ref{eq:model}). The prior on $\mathbf{B}$ is as follows:
\begin{align}
\begin{split}
    B_{k,j} \mid \gamma_{k,j}, \nu^2_B & \sim \gamma_{k,j} \text{Normal}(0,\nu_B^2) + (1 - \gamma_{k,j})\delta_0 \\
    \gamma_{k,j} \mid \theta_{\gamma_j}  & \sim \text{Bernoulli}(\theta_{\gamma_j}) \\
    \theta_{\gamma_j} \mid a_\gamma,b_\gamma & \sim \text{Beta}(a_\gamma,b_\gamma), 
    \end{split}
    \label{eq:B prior}
\end{align}   

\noindent for $j=1,\ldots,p$ and $k=1,\ldots,q$, and with $\delta_0$ a point mass at 0, indicating that when $\gamma_{k,j}$ is 0, $B_{k,j}$ is exactly 0. Here, $\theta_{\gamma_j}$ is the probability of a variable being relevant in $\mathbf{B}_j$. Notice that the mixture prior  \eqref{eq:B prior} allows each variable to be relevant for individual responses  \citep{richardson2010, stingo2010}, as opposed to spike-and-slab constructions that select variables as relevant to either all or none of the responses \citep{brown98}. We also put a non-informative prior on each element of $\mathbf{B}_0$, i.e. $B_{0j} \propto 1$.

\subsection{Prior on precision matrix $\mathbf{\Omega}$}

Next we introduce the prior on the precision matrix $\mathbf{\Omega}$, which allows us to learn a sparse  association network. We consider the prior of  \cite{Wang2015} in the formulation proposed by \cite{Li2017}: 
 \begin{align}
    \begin{split}
    \pi(\mathbf{\Omega} \mid \boldsymbol{\delta},\nu_1,\nu_0,\lambda. \tau) \propto &  \prod_{i<j} \big\{(1 - \delta_{i,j})\text{Normal}(\omega_{i,j} \mid 0,\,\frac{\nu_{0}^2}{\tau}) \text{ + }  
      \delta_{i,j}\text{Normal}(\omega_{i,j} \mid 0,\,\frac{\nu_{1}^2}{\tau})  \big\}\cdot \\
     & \quad \prod_i  \text{Exp}(\omega_{i,i} \mid \lambda/2)\textbf{1}_{\boldsymbol\Omega \in M^+},
     \label{eq:omega_prior}
     \end{split}
\end{align}   

\noindent
where $\nu_0$ and $\nu_1$ are fixed standard deviations, that assume small and large values respectively, $\delta_{i,j}$ is a latent variable indicating whether or not an edge is present between nodes $i$ and $j$, and $\tau$ is a scaling parameter, with a hyperprior $\text{Gamma}(a_\tau,b_\tau)$ that allows to adaptively learn the standard deviations. The original prior of \cite{Wang2015} is obtained by setting $\tau = 1$. Additional complexity can be added to the prior on $\tau$ to include existing knowledge about variable associations, as shown in \cite{Li2017}. The mixture of normals on the off-diagonal precision matrix entries enables the selection of interactions, represented by edges in a network, since non-zero precision matrix entries reflect conditional dependence relationships \citep{Dempster}. Here, entries reflecting conditional independence relations do not equal exactly zero, but get shrunk to close to zero. The diagonal entries are drawn from a common exponential prior. The final term in equation \eqref{eq:omega_prior} expresses a constraint to the space of positive definite matrices $M^+$. This prior is particularly advantageous in our model, as it allows for efficient estimation via the EM algorithm and leads to less bias in graph estimation than the graphical LASSO, as shown by \cite{Li2017}.
 
We complete the modeling of the precision matrix $\mathbf{\Omega}$ by setting the prior on the graph structure, assuming independent Bernoulli distributions on the inclusion of each edge as follows:
\begin{align}
\begin{split}
        \delta_{i,j} \mid \pi & \sim \text{Bernoulli}(\pi) \\ 
        \pi \mid a_\pi, b_\pi & \sim Beta(a_\pi,b_\pi). 
        \label{eq:edge prior}
\end{split}
\end{align}

\noindent

\section{Posterior Inference}
\label{section:comp}

We now discuss how to obtain posterior estimates of the parameters in the model outlined in Section \ref{section:Model}. Instead of a traditional Markov chain Monte Carlo (MCMC) sampler, which can be computationally quite expensive,  we rely on a Variational Inference (VI) approach, which aims to find an approximation of the posterior using optimization methods. VI works by specifying a family of approximate distributions $\mathcal{Q}$, which are densities over latent variables $W$ that are dependent on free parameters $\boldsymbol{\xi}$, and then seeking to find the values of $\boldsymbol{\xi}$ that minimize the Kullback-Leibler (KL) divergence between the approximate distribution and the true posterior. As discussed in \cite{blei2017}, minimizing the KL divergence is equivalent to maximizing the Evidence Lower BOund (ELBO), which is defined as: 
\begin{align}
    \text{ELBO} = \EX_{\xi} [\text{log } p({X , W})] - \EX_{\xi} [\text{log } q(W)],
    \label{eq:ELBO_simple}
\end{align}
with $p({X,W})$ {as} the joint distribution of {the observed data and the latent variables}, and $q(W)$ the {variational distributions of the latent variables}.

The most common approach to obtain an approximating distribution when applying a variational Bayes approach is mean field approximation, where the approximating distribution is assumed to factorize over some partition of the parameters. This is the approach that we adopt for the coefficient vector $\mathbf{B}$. However, a mean field approach for the elements of the precision matrix $\mathbf{\Omega}$ is not appropriate, due to the dependence among the parameters induced by the fact that this matrix is constrained to be symmetric and positive semi-definite. For this reason, the choice of an appropriate approximating distribution for the precision matrix is an open research question. To circumvent this issue, similar to \cite{miao2020}, we adopt a hybrid VI algorithm, with an Expectation-Minimization (EM) step to estimate $\mathbf{\Omega}$ and $\boldsymbol{\delta}$.

Specifically, for $\mathbf{B}$ we use the mean field variational distributions
$q(\mathbf{B},\boldsymbol{\gamma}) = \prod_{j = 1}^{p} \prod_{k = 1}^{q} q(B_{k,j},\gamma_{k,j} \mid \phi_{k,j},\mu_{k,j},\sigma_{k,j}),$
where
$$ 
q(B_{k,j},\gamma_{k,j} \mid \phi_{k,j},\mu_{k,j},\sigma_{k,j}) =
\begin{cases}
\phi_{k,j}\text{Normal}(B_{k,j} \mid \mu_{k,j}, \sigma_{k,j}) \text{ if } \gamma_{k,j} = 1, \\
(1 - \phi_{k,j})\delta_0(B_{k,j}) \text{\hspace{2cm} otherwise,}
\end{cases}
$$
\noindent
with free parameters {$\boldsymbol{\xi}$ =  $\{\phi_{k,j},\mu_{k,j},\sigma_{k,j} \}$.} We then define the ELBO as

\begin{align*}
    \text{ELBO} = & \EX_{\xi} \Bigg[ \text{log} \prod_{j = 1}^{p} \Bigg( p(\mathbf{X}_i \mid \mathbf{Z}_i) p(\mathbf{Z}_i \mid \mathbf{B}_0, \mathbf{M}_i, \mathbf{B}, \mathbf{\Omega})
    \prod_{k = 1}^{q} \Big( p(B_{k,j} \mid \gamma_{k,j}, \nu^2_B) p(\gamma_{k,j} \mid \theta_{\gamma_j}) p(\theta_{\gamma_j} \mid a_\gamma,b_\gamma) \Big) \Bigg) \\
    & p(\mathbf{\Omega} \mid \boldsymbol{\delta},\nu_1,\nu_0,\lambda, \tau) p(\delta_{i,j} \mid \pi) p(\pi \mid a_\pi, b_\pi) \Bigg] - \EX_{\xi} \Bigg[ \text{log} \prod_{j = 1}^{p} \prod_{k = 1}^{q} q(B_{k,j},\gamma_{k,j} \mid \phi_{k,j},\mu_{k,j},\sigma_{k,j}) \Bigg],
\end{align*}
where the first expectation is equivalent to $\EX_{\xi} [\text{log } p(X , W)]$ and the second expectation is $\EX_{\xi} [\text{log } q(W)]$ of equation \eqref{eq:ELBO_simple}, for $W = \{ \mathbf{Z}$, $\mathbf{B}_0$, $\mathbf{B}$, $\mathbf{\Omega}$, $\boldsymbol{\delta}$, $\pi$, $\boldsymbol{\gamma}$, $\boldsymbol{\theta}_\gamma $\}.

The hybrid scheme we use to maximize the ELBO, where the first part is a VI step and the second part is an EM step, is described in detail in the following subsections. In the VI step we update the free parameters, {$\boldsymbol{\xi}$}, by setting the partial derivative of the ELBO with respect to the desired parameters equal to zero. {This minimizes the ELBO with respect to $\boldsymbol{\xi}$. We then further minimize the ELBO by finding the optimal values for the remainder of the latent parameters. For this, we rely on an EM step, by treating $\boldsymbol{\delta}$ as latent parameters and taking the expectation of the ELBO with respect to $\boldsymbol{\delta}$, or equivalently setting $$ Q(\boldsymbol{\theta} \mid \boldsymbol{\theta}^{(t)},\boldsymbol{\xi}^{(t)}) = \EX_{\boldsymbol{\delta}}[ELBO], $$ and optimizing $ Q(\boldsymbol{\theta} \mid \boldsymbol{\theta}^{(t)},\boldsymbol{\xi}^{(t)})$ by finding the maximum a posteriori (MAP) estimate of the remaining parameters $\boldsymbol{\theta}$ = \{$\mathbf{Z}$, $\mathbf{B}_0$, $\mathbf{\Omega}$, $\pi$, $\boldsymbol{\theta}_\gamma$$\}$. The resulting algorithm, which we call SINC (Simultaneous Inference for Networks and Covariates) is described in Algorithm 1. As with traditional EM and VI schemes, parameter updates at each iteration are made with the most current estimates of all other parameters. The algorithm results in MAP estimates for the parameters in $\boldsymbol{\theta}$. Additionally, since no uncertainty about these parameters is used in the updates of the other parameters, the proposed algorithm is only suitable for point estimation. }

\begin{algorithm}[h]
\fontsize{10}{6.5}\selectfont
\caption{SINC Algorithm}
\begin{algorithmic}[1]
\Procedure{Initialize}{$\mathbf{Z},\mathbf{\Sigma}, \mathbf{\Omega}$}       
    \State Set $\mathbf{Z} = \log(\boldsymbol{\alpha} + 1)$ 
    \State $\mathbf{\Sigma}$ = $\tilde{\mathbf{Z}}^T\tilde{\mathbf{Z}}$/N       \Comment{$\tilde{\mathbf{Z}}$ is $\mathbf{Z}$, column centered}
    \State $\mathbf{\Omega} = \mathbf{\Sigma}^{-1}$
\EndProcedure

\While{ELBO has not converged}

\Procedure{VI-Step}{$\mathbf{B}$,$\boldsymbol{\phi}$}
    \For{$j$ in $1:p$}
        \While {$\boldsymbol{\mu}_j^{(t)}$, $\boldsymbol{\sigma}_j^{(t)}$, $\boldsymbol{\phi}_j^{(t)}$ not converged}
        \State Update $\boldsymbol{\mu}_j$ \Comment{Equation (\ref{eq:mu_update})}
        \State Update $\boldsymbol{\sigma}_j$ \Comment{Equation (\ref{eq:sigma_update})}
        \State Update $\boldsymbol{\phi}_j$ \Comment{Equation (\ref{eq:phi update})}
        \EndWhile
    \EndFor
    \State Set $B_{k,j} = \mu_{k,j}\phi_{k,j}$
\EndProcedure

\Procedure{E-Step}{$\boldsymbol{\delta}$}
    \State Update $\mathbf{p^*}$ \Comment{Equation (\ref{eq:p_star update})}
    \State Update $\mathbf{d^*}$ \Comment{Equation (\ref{eq:d_star update})}
\EndProcedure

\Procedure{M-Step}{$\mathbf{B}_0$,$\mathbf{\Omega},\theta_\gamma,\pi,\mathbf{Z}$}
    \For{$j$ in $1:p$}
        \State Update $B_{0j}$ \Comment{Equation (\ref{eq:update B0})}
    \EndFor
    \While {$\mathbf{\Omega}^{(t)}$ not converged}
        \State Column-wise update of $\mathbf{\Omega}$ \Comment{Equations (\ref{eq:omega_ij update},\ref{eq:omega_ii update})}
    \EndWhile
    \For{$j$ in $1:p$}
        \State Update $\boldsymbol{\theta}_{\gamma_j}$ \Comment{Equation (\ref{eq:theta update})}
    \EndFor
    \State Update $\pi$ \Comment{Equation (\ref{eq:pi update})}
    \State Update $\tau$, if applicable \Comment{Equation (\ref{eq:tau update})}
    \State Update $\mathbf{Z}$ \Comment{Maximize Equation (\ref{eq:Z update}) using L-BFGS}
\EndProcedure
\EndWhile
\end{algorithmic}
\end{algorithm}

{Our proposed hybrid algorithm builds upon the similarities between the VI and EM algorithms. As noted in \cite{blei2017}, the first term of equation \eqref{eq:ELBO_simple} is the expected complete log likelihood, which is optimized by the EM algorithm. Since no variational distributions are proposed for the parameters in $\boldsymbol{\theta}$, updating those parameters is achieved by optimizing $\log p(W,Y)$ in equation \eqref{eq:ELBO_simple}. As an alternative perspective to highlight the similarity, we could say that we have assigned a point mass as our variational distribution for these latent parameters. Optimizing the ELBO would then lead to the same result, since taking the partial derivative of the ELBO with respect to the variables with point mass variational distributions would result in optimizing $\EX_{\xi}\text{log}[p(W,Y)]$. By stating the algorithm in this way, we can interpret our approach as a proper VI scheme, solved via an EM step similar to \cite{Titsias2011}, which affords us the confidence of VI guarantees of previous literature \citep{blei2017}.}

\subsection{VI Step}
Here we use a Variational Inference step to estimate the regression coefficients $\mathbf{B}$ by updating the free parameters $\mu_{k,j}$, $\sigma_{k,j}$, and $\phi_{k,j}$, where $\mu_{k,j}$ and $\sigma_{k,j}$ are the mean and variance, respectively, of $B_{k,j}$ when $\gamma_{k,j} = 1$, and $\phi_{k,j}$ is the probability $\gamma_{k,j} = 1$ resulting in $B_{k,j} \neq 0$. Following the work of \cite{Titsias2011} and \cite{carbonetto} the free parameters can then be updated as

\begin{align}
    \mu_{k,j} & =  \frac{\sigma_{k,j}}{\sigma_{j}^*} \Big[\big\{\mathbf{M}^T(\mathbf{Z}_j - B_{0j})\big\}_k - \sum_{l \neq k} (\mathbf{M}^T\mathbf{M})_{lk}\phi_{l, k} \mu_{l,k} \Big],
    \label{eq:mu_update}  \\
    \sigma_{k,j} & = \frac{\sigma_{j}^*}{(\mathbf{M}^T\mathbf{M})_{k,k} + 1/\nu_B}. \label{eq:sigma_update} \\
    \phi_{k,j} & = \text{Logit}^{-1}\big(\text{log}\frac{\theta_{\gamma_j}}{1 - \theta_{\gamma_j}} - \frac{1}{2} \text{log} \frac{ \sigma_{j,k}}{\nu_B \sigma^*_j} + \frac{\mu_{i,j}^2}{2\sigma_{j,k}^2} \big) \label{eq:phi update},
\end{align}

\noindent
which is interpreted as the probability that $B_{k,j}$ comes from the continuous distribution, while (1 - $\phi_{k,j}$) is the probability that $B_{k,j}$ comes from the point mass at 0. Updating each column of $\mathbf{B}$ can then be done independently and, when resources are available, these updates can be done in parallel. While updating a column of $\mathbf{B}_j$, each component of $\boldsymbol{\mu}_{j}$, $\boldsymbol{\sigma}_{j}$, $\boldsymbol{\phi}_{j}$ is updated given all other components. This component-wise update of $\boldsymbol{\mu}_{j}$, $\boldsymbol{\sigma}_{j}$, $\boldsymbol{\phi}_{j}$ is repeated until ELBO($\boldsymbol{\mu}_j, \boldsymbol{\sigma}_j, \boldsymbol{\phi}_j$) has converged. Once all $\boldsymbol{\mu}$ and $\boldsymbol{\phi}$ have been updated, the individual elements of $\mathbf{B}$ are assigned $\EX (B_{k,j}) = \mu_{k,j}\phi_{k,j}$. Once the SINC algorithm has converged, as common in variational spike-and-slab literature \citep{carbonetto,Huang2016,miao}, we set ${B}_{k,j} = \mu_{k,j} \text{ if }\phi_{k,j} > 0.5 \text{ and } {B}_{k,j} = 0  \text{ if } \phi_{k,j} \leq 0.5$. The threshold of 0.5 is equivalent to selecting the median model of \cite{Barbieri2004} and can be adjusted to include or exclude more covariates, but the threshold of 0.5 is the most commonly used.

\subsection{E Step}
In this step, we focus on updates to the edge inclusion parameter $\delta_{i,j}$. For the first step we take the expectation of the posterior distribution, treating $\boldsymbol{\delta}$ as the latent variable. We define $Q(\mathbf{\Theta} \mid \mathbf{\Theta}^{(t)})$ as $ E_{\boldsymbol{\delta} \mid \boldsymbol{\Omega}^{(t)} , \pi, \textbf{X}} \big( \text{log } p(\mathbf{Z}, \mathbf{\Omega}, \mathbf{B}, \mathbf{B}_0, \pi \mid \mathbf{X}, \mathbf{M}) \mid \boldsymbol{\Omega}^{(t)} , \pi^{{(t)}}, \textbf{X} \big).$ Following the results shown in \cite{Li2017}, the E step can be broken into two steps:
\begin{align}
E_{\boldsymbol{\delta} \mid \boldsymbol{\Omega^{({t})}} , \pi, \textbf{X}} [\delta_{i,j}] & = p^{*}_{ij} \equiv \frac{a_{ij}}{a_{ij} + b_{ij}} \label{eq:p_star update} \\
E_{\boldsymbol{\delta} \mid \boldsymbol{\Omega^{({t})}} , \pi , \textbf{X}} \Big[\frac{1}{\nu_0^2 \tau^{-1}(1 - \delta_{i,j}) + \nu_1^2 \tau^{-1}\delta_{i,j}}\Big] & = d^{*}_{ij} \equiv \Bigg( \frac{1 - p^{*}_{ij}}{\nu_0^2} + \frac{p^{*}_{ij}}{\nu_1^2} \Bigg) \tau^{-1}, \label{eq:d_star update}
\end{align}
\vskip.2cm \noindent
where $a_{ij} = p(\omega_{i,j} \mid \delta_{i,j} = 1) \pi$ and $b_{ij} = p(\omega_{i,j} \mid \delta_{i,j} = 0) (1 - \pi)$, and $(i,j)$ is the $(i,j)^{th}$ entry of the precision matrix, where $i$ and $j$ $\in$ $\{1, \ldots, P\}$.

\subsection{M Step}
The remainder of the unknown parameters can be found by maximizing the posterior distribution with regards to each of the parameters we are interested in. Here, we first update the column-wise centering parameters $B_{0j}$ independently as
\begin{align}
    B_{0j} = \frac{\sum_{i = 1}^N Z_{i,j} - \mathbf{M}_i\mathbf{B}_j}{N}.
    \label{eq:update B0}
\end{align}

\noindent Next, we update the precision matrix, $\mathbf{\Omega}$. Following \cite{Li2017} and \cite{Wang2015} the conditional distribution of each column of $\mathbf{\Omega}$ can be found in closed form. For this, let 

\begin{center}
$ \mathbf{\Omega} $ = $\Bigg($ \begin{tabular}{c c c c}
$\mathbf{\Omega}_{11}$ & $\boldsymbol{\omega}_{12}$ \\
$\boldsymbol{\omega}^T_{21}$ & $\omega_{22}$ 
\end{tabular} $\Bigg)$,
$(\mathbf{Z} - (\mathbf{MB} + \mathbf{B}_0))^T(\mathbf{Z} - (\mathbf{MB} + \mathbf{B}_0))$ = $\Bigg($ \begin{tabular}{c c c c}
$\mathbf{S}_{11}$ & $\mathbf{s}_{12}$ \\
$\mathbf{s}_{21}$ & $s_{22}$ 
\end{tabular} $\Bigg)$. 
\end{center} 
\noindent 
Then, the conditional distributions are
\begin{align}
\begin{split}
\boldsymbol{\omega}_{12} & \sim \text{Normal}(-\mathbf{C}\mathbf{s}_{12} , \mathbf{C}), \\
\omega_{22} - \mathbf{\omega}_{12}\mathbf{\Omega}_{11}\mathbf{\omega}_{12} & \sim \text{Gamma}(1 + \frac{n}{2}, \frac{\lambda + s_{22}}{2}), \\
\mathbf{C} & = ((s_{22} + \lambda)\mathbf{\Omega}^{-1} + \text{diag}(\nu_{\boldsymbol{\delta}_{12}}))^{-1}.
\label{eq:omega_ij update}
\end{split}
\end{align}
We can then do a column-by-column update as 
\begin{align}
\begin{split}
\boldsymbol{\omega}_{12}^{k+1} & = -((s_{22} + \lambda)(\mathbf{\Omega}^{k+1})^{-1} + \text{diag}(d^{*}_{ij}))^{-1})\mathbf{s}_{12}, \\
\omega_{22}^{k+1} & = \boldsymbol{\omega}_{12}^{k+1}\mathbf{\Omega}_{11}^{k+1}\boldsymbol{\omega}_{12}^{k+1} + \frac{n}{\lambda + s_{22}}.
\label{eq:omega_ii update}
\end{split}
\end{align}

\noindent
The point estimates of $\theta_\gamma$ and $\pi$ are also updated as
\begin{align}
    \theta_\gamma & = \frac{\sum \phi_{i,j} + a_\gamma - 1}{p + a_\gamma + b_\gamma -2} \label{eq:theta update}, \\
    \pi & = (a_\pi + \sum_{i<j} \delta_{ij} - 1) / (a_{\pi} + b_{\pi} + \frac{p(p-1)}{2} - 2).
    \label{eq:pi update}
\end{align}
If using the adaptive scale parameter, $\tau$, an additional update is done by setting
\begin{align}
    \tau = \frac{a_\tau - 1 + 0.5(P \times (P - 1))}{b_\tau - 2 + 0.5\sum_{i < j} \omega_{i,j}^2 d^*_{i,j}}.
    \label{eq:tau update}
\end{align}

\noindent
Finally, the matrix of latent variables can be estimated by finding a point estimate for each entry of the matrix. This is done by updating each row of the matrix independent of the others. As shown in \cite{mLDM}, the objective function to optimize with respect to $\mathbf{Z}$ is
\begin{align}
    &\text{log }P(\mathbf{Z} \mid \mathbf{X}, \mathbf{M}, \mathbf{B}_0, \mathbf{B}, \mathbf{\Omega}) =  -\frac{1}{n}\sum_{i = 1}^{n}\Bigg[\sum_{j = 1}^{p} \tilde{\Gamma}(\alpha_{ij} + x_{ij}) - \tilde{\Gamma}(s(\boldsymbol{\alpha}_{i}) + s(\mathbf{X}_{i})) - \sum_{j = 1}^{p} \tilde{\Gamma}(\alpha_{ij}) + \tilde{\Gamma}(s(\boldsymbol{\alpha}_{i}))\Bigg] - \notag \\ 
    & \qquad \qquad \qquad \qquad \qquad \qquad \quad \frac{1}{2} \text{ log}\mid \mathbf{\Omega} \mid + \frac{1}{2n} \sum_{i = 1}^{n} (\mathbf{Z}_i - (\mathbf{B}_0 + \mathbf{M}_i\mathbf{B})) \mathbf{\Omega} (\mathbf{Z}_i - (\mathbf{B}_0 + \mathbf{M}_i\mathbf{B})), \label{eq:Z update}
\end{align}
{where $\tilde{\Gamma}$ is the log-gamma function, and $s(x_i)$ =  $\sum_{j = 1}^{p} x_{ij}$.}
To accomplish optimization of each $\mathbf{Z}_i$ we use the limited-memory quasi-Newton (L-BFGS) algorithm, which is a quasi-newton gradient descent method that makes use of the inverse gradient to direct where to search through the variable space.

For posterior inference, we iterate through the VI step, which iterates between updating $\phi_{k,j}$, $\mu_{k,j}$, and $\sigma_{k,j}$, and the E and M steps, which updates $\mathbf{\Omega}$ one column at a time, until the algorithm has converged. For both the VI and the M steps, we run each of those steps until the respective parameter estimates have converged. We determine the algorithm to have converged if the ELBO changes by less than a predefined tolerance from one iteration to the next. To obtain a selected network and set of covariates based on these posterior estimates, we select edges $i,j$ with $p_{i,j}^*$ in equation \eqref{eq:p_star update} $\geq$ 0.50, and covariate associations with $\phi_{k,j}^*$ in equation \eqref{eq:phi update} $\geq$ 0.50. In practice, both the $p^*_{j,k}$ and $\phi_{k,j}$ values, which reflect the posterior probabilities for the selection of edges and covariates, tend to converge to values close to 0 or close to 1. A similar trend has been noted by \cite{Kook2020}, who reported that the variational parameters for the marginal posterior probabilities of inclusion tended to become more widely separated as the algorithm converges.

\section{Simulation Study}
\label{section:sim}
We now compare the performance of our method to existing approaches in a simulation setting designed to mimic the application to microbiome data described later in the paper.

\subsection{Simulation Setup}
Simulated data were generated with the following steps. First the covariates, $\mathbf{M}$, were generated from a normal distribution $\text{MVNorm}(\mathbf{0},\mathbf{I}_p)$, and subsequently scaled. The values of the regression coefficients $\mathbf{B}$, related to $\textbf{M}$, were then sampled. Each element of $\textbf{B}$, $B_{k,j}$, was assigned either a random value between $[-1,-0.5]$ with probability 0.1, a value in $[0.5,1]$ with probability 0.1, or else 0. Each $B_{0j}$ was then sampled from the interval $[6,8]$ with probability 0.2, and from $[2,4]$ with probability 0.8. This allowed for some variables to have larger counts and others to be sparser, as common in microbiome data. Simulated counts were then sampled by first drawing $\textbf{Z}$ from $\text{MVNorm}(\mathbf{MB} + \mathbf{B}_0, \mathbf{\Omega})$, with $\mathbf{\Omega}$ as described below, and then assigning $\boldsymbol{\alpha}$ as exp($\textbf{Z}$). Finally, $\mathbf{h}_i$ was sampled as a random draw from Dirichlet($\boldsymbol{\alpha}_i$), and $\mathbf{X}_i$ drawn from a Multinomial($\mathbf{h}_i$, $\text{nint}(n_i)$), with $n_i$ generated from a $\text{Normal}(3000,250)$, allowing for samples to have different numbers of total counts, and nint() the nearest integer function. We set $p = 100$, $q = 50$ and $n = 300$.

\begin{figure}[h]
\centering
    \includegraphics[width=1.55in,height=1.55in,page = 1]{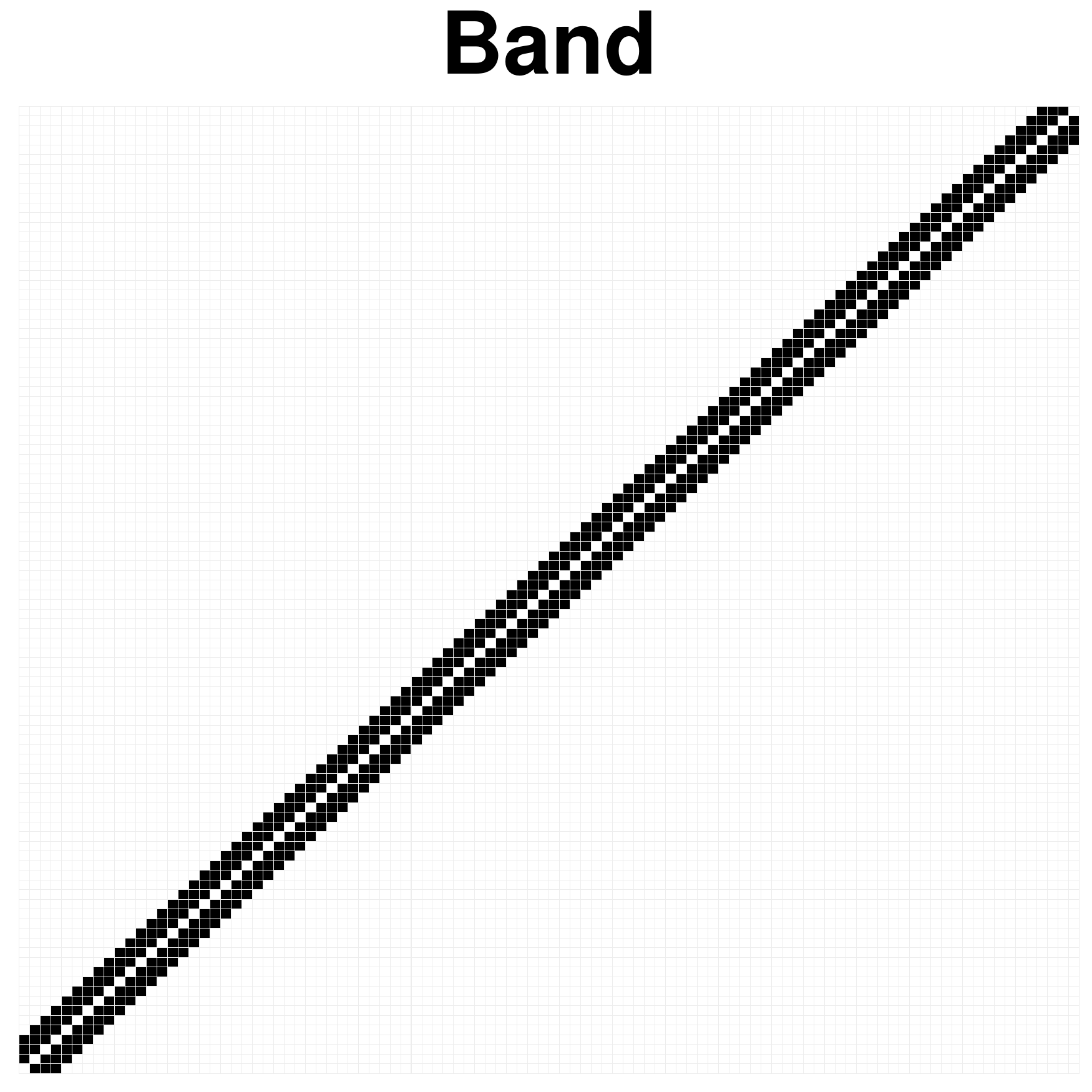}
    \includegraphics[width=1.550in,height=1.55in,page = 1]{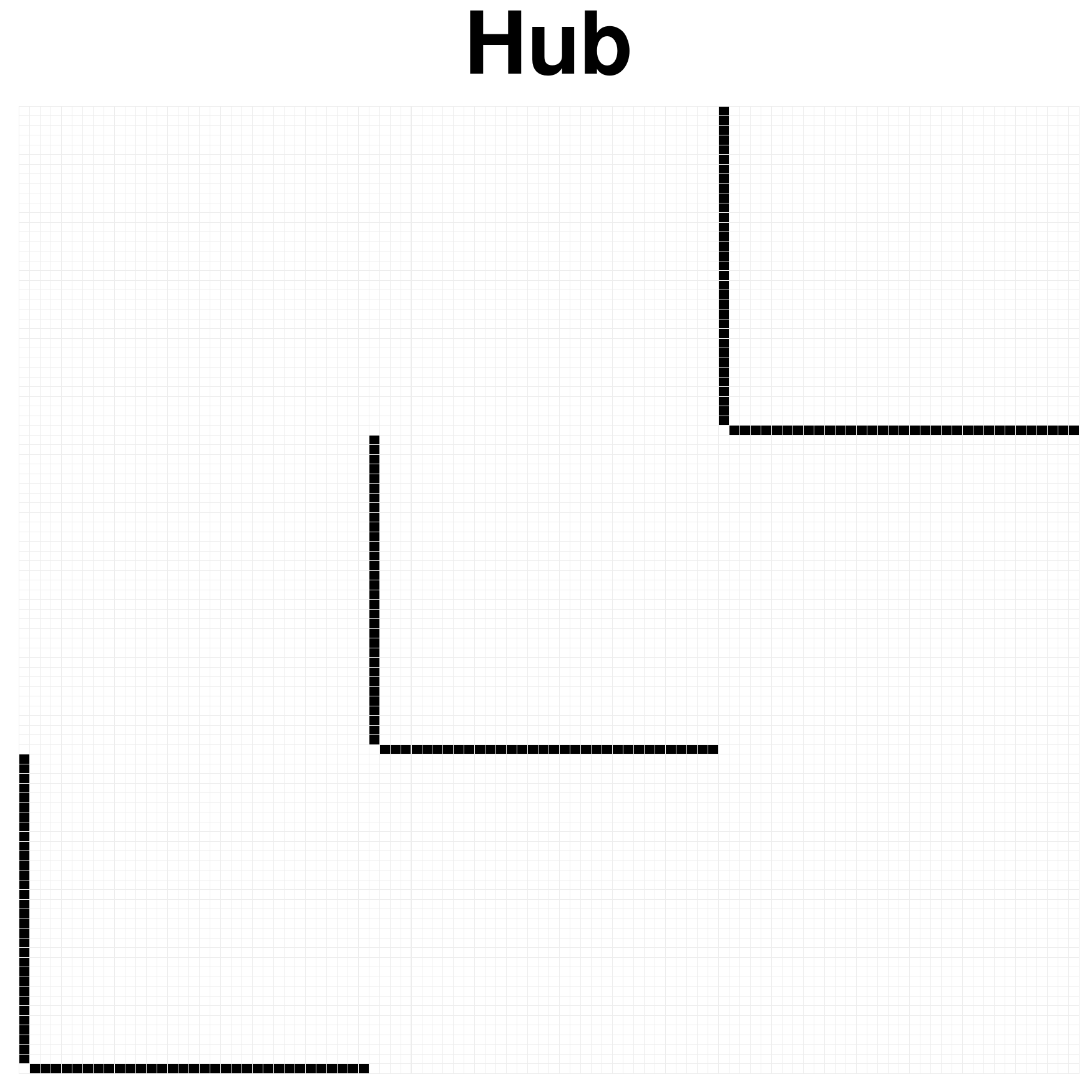} 
    \includegraphics[width=1.550in,height=1.55in,page = 1]{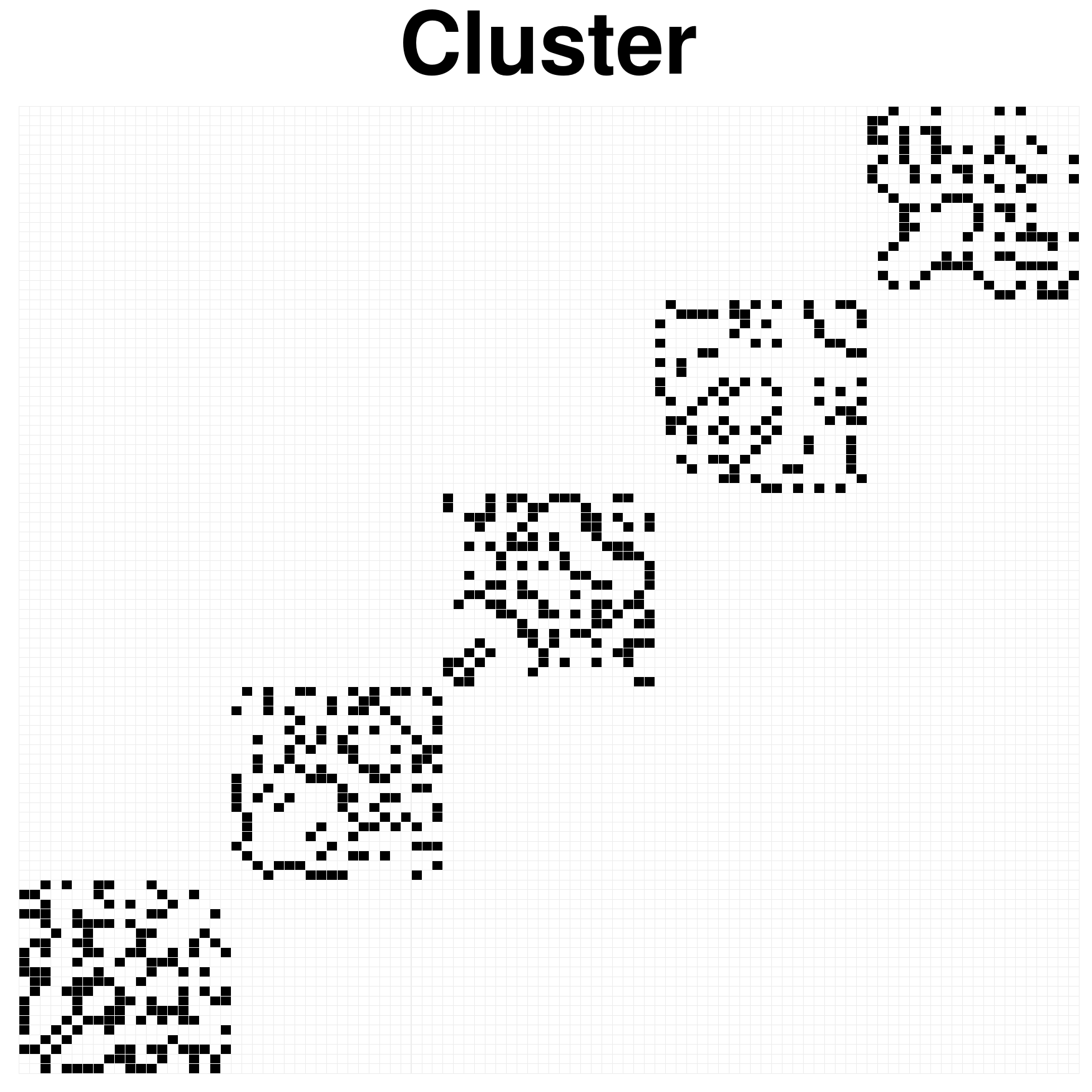}
    \includegraphics[width=1.550in,height=1.55in,page = 1]{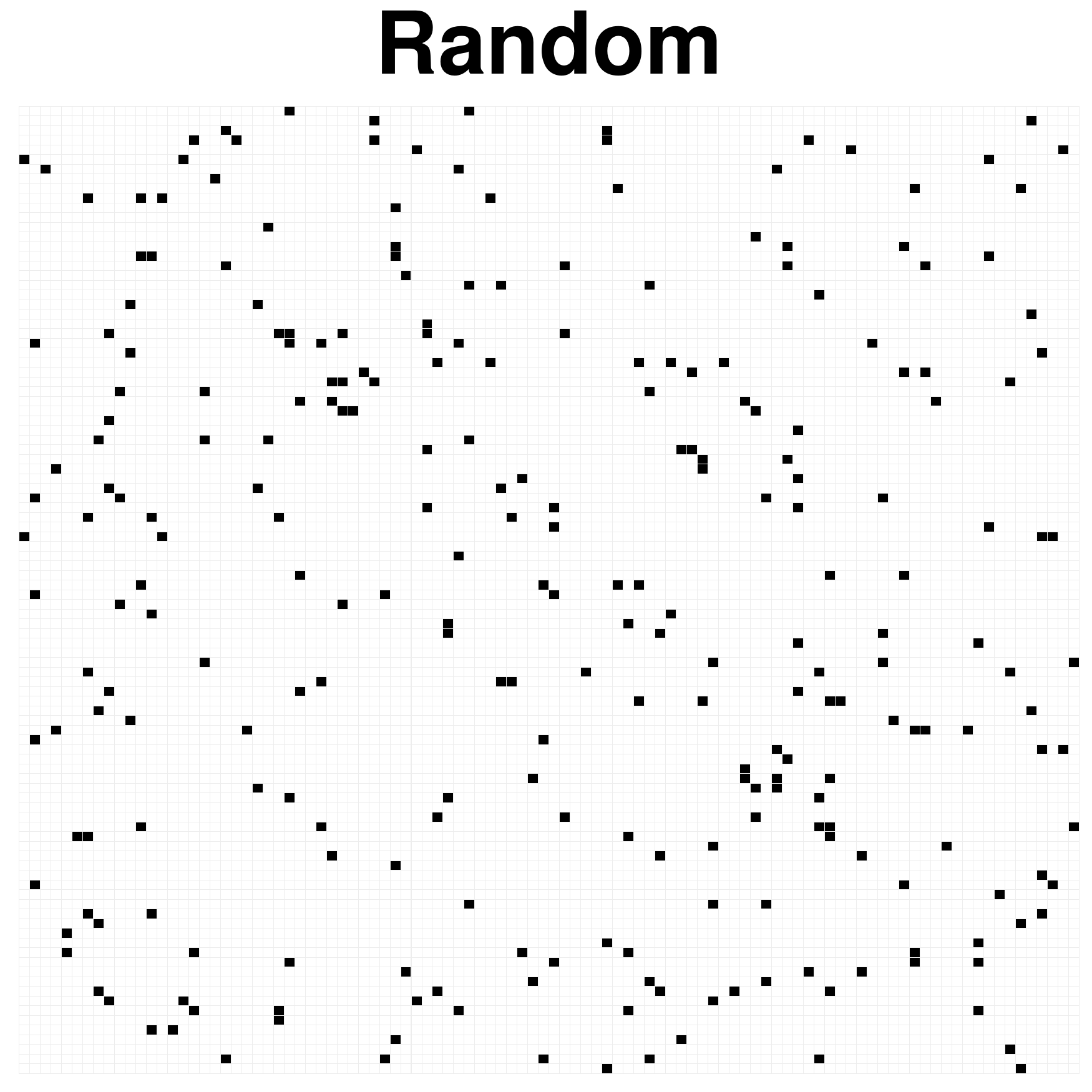}
    \caption{Simulation study: Example networks used in the simulation studies. Black boxes represent true edges, and light gray boxes correspond to no edge. For the cluster and random graphs, the actual networks that generated the data were different for each simulated data set, but each simulation kept the blocked shape or random network, respectively.}
\label{fig:examplegraphs}
\end{figure}

To explore performance for a range of possible  network structures, we simulated a variety of configurations. These networks were created using the R package \texttt{huge} \citep{R_huge}. For this simulation, we used a band, cluster, hub, and random graph structure. An example of what these networks look like can be seen in Figure \ref{fig:examplegraphs}. Band graphs and random graphs are common test cases for network learning, while the hub and cluster graphs capture some aspects of biological networks, such as highly connected nodes and  community structure \citep{Girvan2002}. The probability of an edge in the network was set to 0.025 for the random graph and 0.30 for each cluster in the cluster graph. The bandwidth in the graph was set to 3 and the number of hubs in the hub graph was set to 3. The precision matrix $\mathbf{\Omega}$ used to generate the simulated data was also constructed using the function \texttt{huge.generate} of the R package \texttt{huge} \cite{R_huge}. Parameters \texttt{v} and \texttt{u} of \texttt{huge.generate}, which control the off diagonal elements of the precision matrix and magnitude of the partial correlations, were set to 1 and 0.0001, respectively.

The results we report below for our proposed model were obtained with the following hyperparameter settings.  Fairly non-informative priors were set by choosing $a_\gamma=b_\gamma=2$ in the prior probability of inclusion \eqref{eq:B prior} of each covariate. The same setting was used for the hyperparameters $a_\pi$ and $b_\pi$ in prior \eqref{eq:edge prior}, which determines the prior probability of an edge being included. The standard deviation of the prior on the selected regression coefficients, $\nu_B$ in equation \eqref{eq:B prior}, was set to 1.  Following guidelines given by \cite{Wang2015} and \cite{Li2017}, we set $\nu_1=10$, the standard deviation in prior \eqref{eq:omega_prior} on the off-diagonal precision matrix entries corresponding to selected edges, and fit the model across a grid of values, ranging from 0.0001 to 0.1, of $\nu_0$, the prior standard deviation of off-diagonal elements of the precision matrix corresponding to non selected edges. We then chose the final model by using the $\nu_0$ value that gave sparsity closest to 0.10. This sparsity level was selected arbitrarily, and did not result in any specific advantage for our method, as none of the simulation networks had sparsity equal to 0.10. Finally, the rate parameter $\lambda$ of equation \eqref{eq:omega_prior}, which appears in the prior on diagonal elements of the precision matrix, was set to 150. We also compare the model when the scaling parameter, $\tau$, is learned, and use a Gamma(2,2) prior to do so. We comment on the sensitivity of the results to parameter choices in Section \ref{ssec:influence} below.

\subsection{Simulation Results}
We compare the performance of our method to several existing alternative approaches in terms of accuracy in network estimation and covariate selection. For comparison, we used mLDM \citep{mLDM}, which is specifically designed for estimating networks of compositional data while controlling for covariates, and mSSL-DPE \citep{Deshpande2019}, which we applied to the centered log ratio (CLR) transformed of the simulated count data, a common method to account for the compositionality \citep{gCoda,spiecEasi}. For network estimation, we also considered SpiecEasi \citep{spiecEasi}, which applies graphical LASSO to the CLR-transformed data. These methods were applied by using the default selection criteria in their respective R packages. To more precisely characterize factors contributing to the network estimation performance of the SINC method, we apply a version of SINC with the covariate effects $\mathbf{B}$ constrained to be $\mathbf{0}$. A comparison of the results from this constrained version of SINC to those of SpiecEasi reflects the performance advantages arising from differences in the network estimation procedure, while comparison to the full unconstrained SINC method provides quantitative insight on the benefit of simultaneous estimation of covariate effects on network recovery. Similarly, for the comparison of variable selection accuracy, we apply  a constrained version of SINC with $\mathbf{\Omega}$ fixed to the identity matrix. Comparison of the results from this approach to those of the full unconstrained SINC method illustrates the added value of accounting for the residual covariance in estimation of $\mathbf{B}$. 

\begin{table}[!h]
\begin{center}
{\small \renewcommand{\arraystretch}{0.75}
\vskip.2cm
\textbf{\large{Network estimation accuracy}} \vskip.2cm
\centering
\begin{tabu}{|l|lllll||lllll|}
\hline
         & TPR            & FPR            & F1             & MCC            & AUC            & TPR            & FPR            & F1             & MCC            & AUC            \\
          \hline \hline \hline
          & \multicolumn{5}{c||}{Random} &
                \multicolumn{5}{c|}{Hub}        \\
          \hline
mSSL-DPE   & 0.000          & 0.015          & 0.000          & -0.019          & 0.499          &        0.001 & 0.024 &  0.001 & -0.021 & 0.450         \\
SpiecEasi & 0.013          & \textbf{0.009}          & 0.018  &  0.005         &  0.563 &       0.011 & \textbf{0.009} & 0.015 & 0.003 & 0.528    \\
mLDM     & 0.277 & 0.181 & 0.067 & 0.039 & 0.560 & \textbf{0.440}        &    0.248       &   0.080        &   0.069 & 0.602          \\
\rowfont{}
SINC (\textbf{B} = 0)  & {0.276} & {0.225} & {0.056} & {0.020} & {0.542} & {0.253} & {0.241} & {0.038} & {0.004} & {0.507}
\\
\rowfont{}
\textcolor{black}{SINC ($\tau=1$)}     & {0.420} & 0.093          & {0.167}          & {0.169}          & {0.750} & {0.175} &   0.096        & {0.059} &  {0.037} & {0.613} \\
\rowfont{}
SINC ($\tau$ learned) & \textbf{0.598} & 0.091 & \textbf{0.237} & \textbf{0.263} & \textbf{0.838} & 0.294 & 0.094 & \textbf{0.098} & \textbf{0.094} & \textbf{0.689}\\
\hline \hline
            & \multicolumn{5}{c||}{Cluster} &
                \multicolumn{5}{c|}{Band}        \\
          \hline
mSSL-DPE   &   0.005 &  0.0181 &  0.008 & -0.0230   & 0.446      &    0.003 & 0.022  & 0.004 & -0.032  &   0.468      \\
SpiecEasi & 0.013 &		\textbf{0.010}	&	0.0182	& 0.005 & 0.563
         & 0.012 & \textbf{0.009} & 0.020 & 0.007 & 0.544          \\
mLDM      & 0.232           & 0.155  & 0.126  & 0.051       &  0.544         & 0.440         & 0.248 & 0.080 & 0.069  & 0.602        \\
\rowfont{}
{SINC (\textbf{B} = 0)}  & {0.272} & {0.229} & {0.110} & {0.024} & {0.534} & {0.294} & {0.242} & {0.114} & {0.028} & {0.533}
\\
\rowfont{}
\textcolor{black}{SINC ($\tau=1$)}     & {0.294} &      0.084     &    {0.223}       &    {0.169}       & {0.678} & {0.311} &   0.088        & {0.230} & {0.175} & {0.685} \\
\rowfont{} 
SINC ($\tau$ learned) & \textbf{0.411} & 0.080 & \textbf{0.306} & \textbf{0.261} & \textbf{0.741} & \textbf{0.446} & 0.089 & \textbf{0.312} & \textbf{0.269} & \textbf{0.737}\\
\hline
\end{tabu}
\caption{Simulation results for network selection. mSSL-DPE refers to the method of \cite{Deshpande2019}, SpiecEasi to the method of \cite{spiecEasi}, mLDM to \cite{mLDM}, SINC (\textbf{B} = 0) to the modified version of the proposed model with the covariate estimates fixed, and SINC to the proposed model. Random, Hub, Cluster, and Band refer to the underlying shape of the network, as illustrated in Figure \ref{fig:examplegraphs}}
\label{table:Network Estimates}
}
\end{center}
\end{table}

\begin{table}[!h]
\begin{center}
{\small \renewcommand{\arraystretch}{0.75}
\textbf{\large{Variable selection accuracy}} \vskip.2cm
\centering
\begin{tabular}{|l|llll||llll|} 
\hline
        & TPR            & FPR            & F1             & MCC                        & TPR            & FPR            & F1             & MCC                        \\
          \hline \hline \hline
             & \multicolumn{4}{c||}{Random} &
                \multicolumn{4}{c|}{Hub}        \\
          \hline
mSSL-DPE   & 0.840         & \textbf{0.000}          & 0.912          & 0.898          &        0.868 & \textbf{0.000} & 0.929 & 0.915                    \\
mLDM      & 0.609         & 0.003  &  0.751       & 0.734     & 0.612         & 0.003 &  0.754 &  0.738        \\
SINC ($\mathbf{\Omega} = \textbf{I}$)   & 0.808 & 0.003 & 0.871 & 0.889 & 0.813 & 0.001 & 0.887 & 0.894
\\
 SINC ($\tau=1$)     &  0.914 &  0.001       &    0.943       & 0.953          &  0.925 &  0.000 & \textbf{0.952}        & \textbf{0.960}   \\
SINC ($\tau$ learned)    & \textbf{0.917} &    0.000       &    \textbf{0.947}       & \textbf{0.956}          &  \textbf{0.926} &0.001 & \textbf{0.952}        & \textbf{0.960}   \\
\hline \hline
             & \multicolumn{4}{c||}{Cluster} &
                \multicolumn{4}{c|}{Band}        \\
          \hline
mSSL-DPE              &  0.837 & \textbf{0.000} & 0.910 & 0.900         &   0.853 & \textbf{0.000} & 0.920 & 0.906       \\
mLDM                 & 0.605 & 0.003  & 0.747       & 0.730          &     0.597     & 0.003 & 0.741 &  0.725         \\
SINC ($\mathbf{\Omega} = \textbf{I}$)  & {0.791} & {0.006} & {0.854} &{0.873} & {0.808} & {0.000} & {0.887} & {0.893}
\\
SINC ($\tau=1$)     & 0.908 & 0.000       &  0.941        &   0.951       &   0.921 & 0.001 &  0.947         &  0.957   \\
SINC ($\tau$ learned)   & \textbf{0.910} &  0.000       &    \textbf{0.942}       &  \textbf{0.952}          & \textbf{0.922} & 0.000 &  \textbf{0.950}        &\textbf{0.959}   \\
\hline
\end{tabular}
\caption{Simulation results for covariate selection. mSSL-DPE refers to the method of \cite{Deshpande2019}, mLDM to the method of \cite{mLDM}, SINC ($\mathbf{\Omega} = \textbf{I}$) to the modified version of the proposed model with the precision matrix fixed, and SINC to the proposed model. SpiecEasi is omitted from the comparison, as it does not perform selection or adjustment for covariates. Random, Hub, Cluster, and Band refer to the underlying shape of the network, as illustrated in Figure \ref{fig:examplegraphs}}
\label{table:B estimates}
}
\end{center}
\end{table}

We report results in terms of true positive rate (TPR), false positive rate (FPR), F1 score, and Matthew's correlation coefficient (MCC). For edge selection, we also report the  area under the curve (AUC). This was calculated, for the SINC method, over a grid of $\nu_0$ values, and for the mLDM and mSSL-DPE methods by using the LASSO penalization parameter for the coefficients associated with the best selected graph, and then varying the graph penalization parameter over a grid of values. The AUC for SpiecEasi was calculated by varying the penalization parameter. Tables \ref{table:Network Estimates} and \ref{table:B estimates} show the results for network estimation and variable selection, respectively. From Table \ref{table:Network Estimates}, we can see that mSSL-DPE and SpiecEasi are generally not competitive in terms of their performance, with low F1, MCC, and AUC values. This is likely because mSSL-DPE was not designed for compositional data, and SpiecEasi is not able to account for the effects of the covariates on the counts. We also see that mLDM performs better than the other two methods but is still outperformed by the proposed model, which does better in all F1, MCC, and AUC scores across all of the network structures except Hub. Finally, we see that when the proposed model does not control for additional covariates, the network estimation scores decrease and are comparable to the other methods. Across all methods, SINC while learning $\tau$ performed best in all network structures in terms of F1, MCC, and AUC. The performance metrics for SINC are pretty similar across all network types, though the best performance is achieved on the random graph, while the hub and cluster settings are more challenging. From Table \ref{table:B estimates} we can see that mSSL-DPE performs quite well in selecting the covariates of interest. In fact, its performance is very close to the proposed model, SINC, on all metrics. Similarly, the proposed model while holding the estimated network and precision matrix fixed performs well for coefficient estimation, but does not do as well as mSSL-DPE and the full version of the proposed model. Additionally, we did not see any significant difference in performance in the full SINC models when $\tau$ is fixed or learned. SpiecEasi is not included in Table \ref{table:B estimates}, as it is not able to select or adjust for relevant covariates, which is a limitation of the method. 

We did not compare our model to MCMC approaches because of the computational complexity resulting from a lack of conjugacy. We did, however, experiment by using a Monte Carlo draw to update $\mathbf{\Omega}$ at each iteration of SINC, and found that the point estimates of SINC without a Monte Carlo step were very close to the mean of the Monte Carlo draws.

\subsection{Influence of parameters}
\label{ssec:influence}
\begin{figure}[!h]
\centering
    \includegraphics[width=2.5in,height=2.2in]{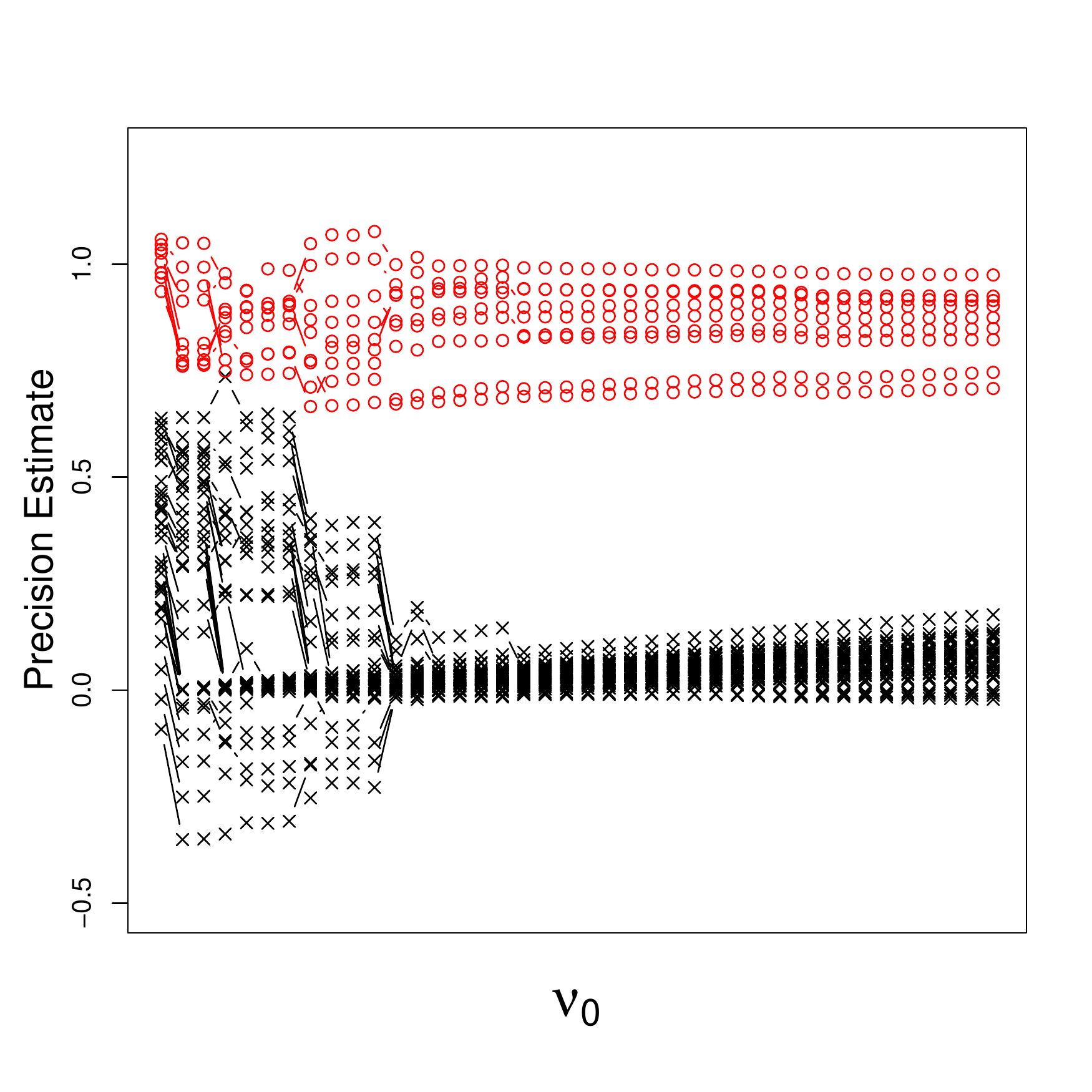}
    \includegraphics[width=2.5in,height=2.2in]{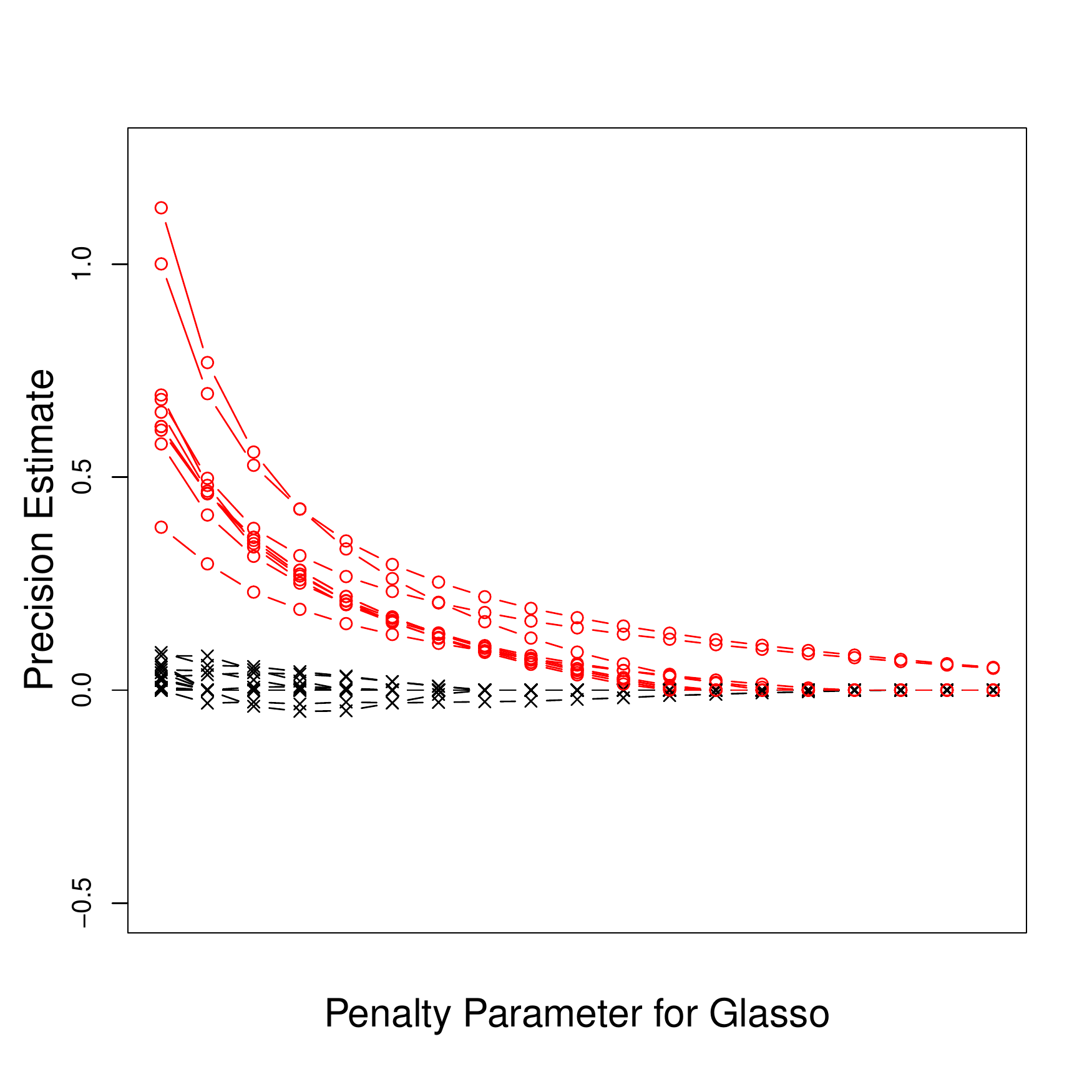} \\
    \includegraphics[width=2.5in,height=2.2in]{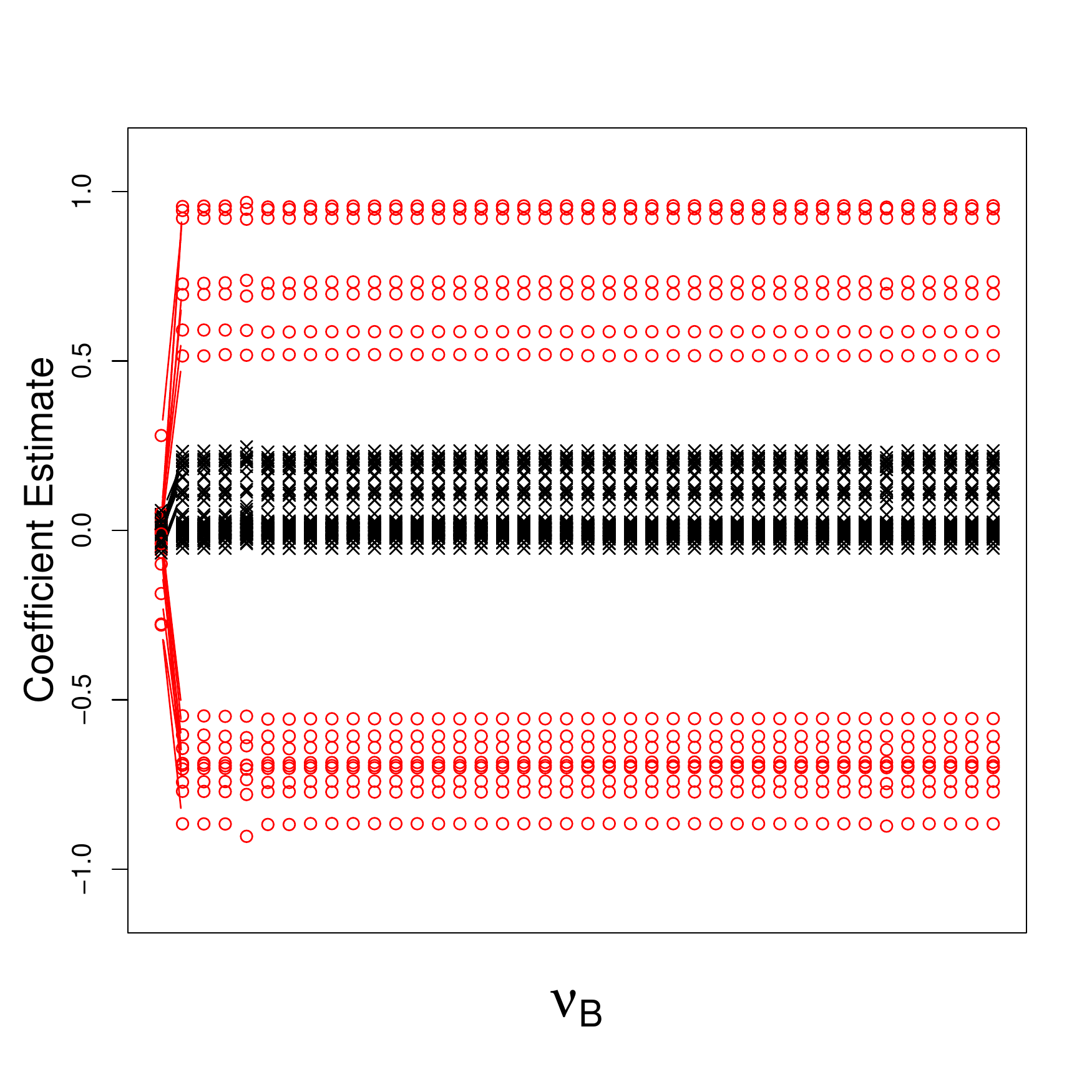}
    \includegraphics[width=2.5in,height=2.2in]{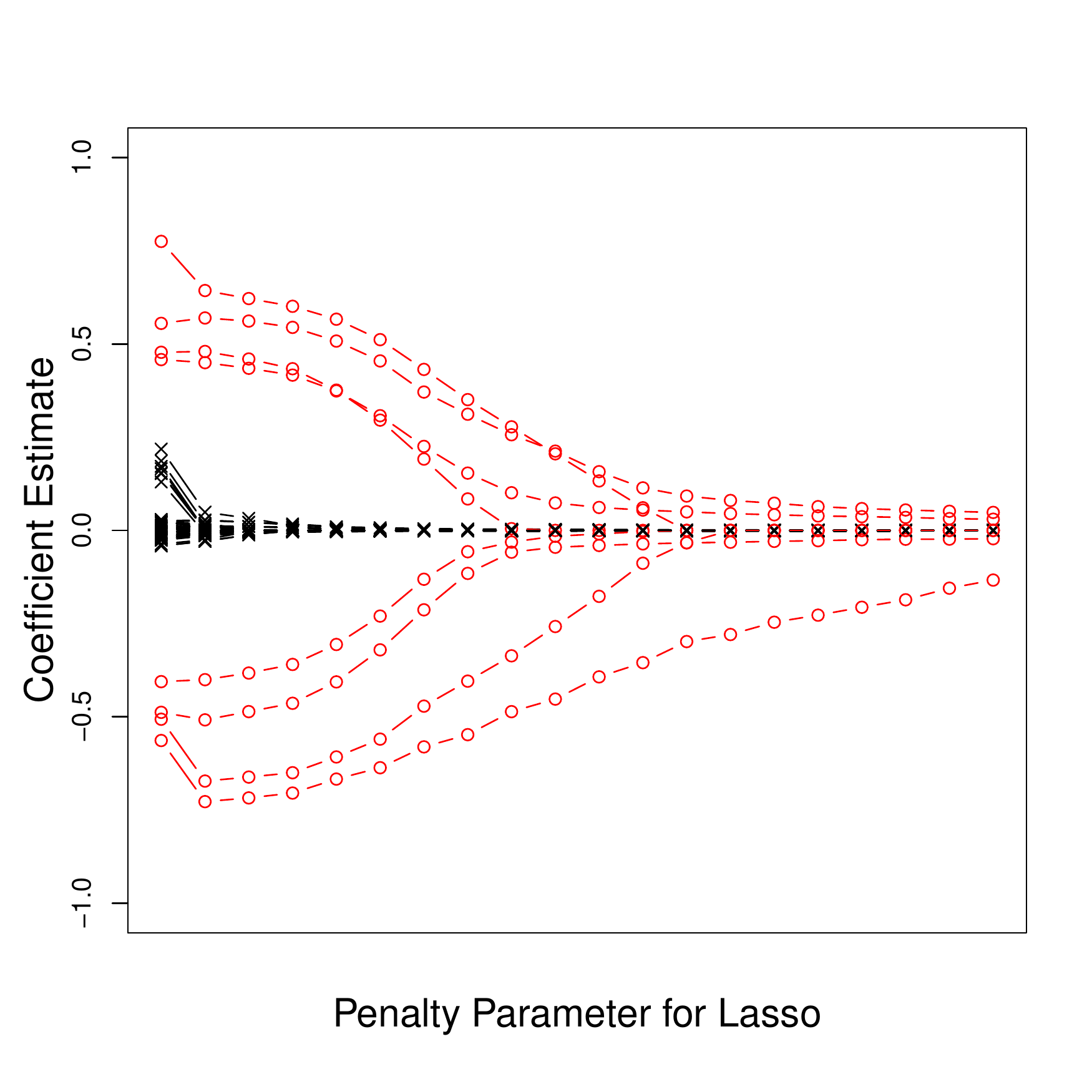}
    \caption{Simulation study: The top-left plot shows the estimated $\mathbf{B}$ coefficients by the proposed model when increasing the variance parameter $\nu_B$ of the spike-and-slab prior \eqref{eq:B prior}. The top-right plot shows the estimated $\mathbf{B}$ coefficients via mLDM when increasing the LASSO penalty parameter. The bottom-left plot shows the estimated off-diagonal values of the precision matrix $\boldsymbol{\Omega}$ when increasing the variance parameter $\nu_0$ in (\ref{eq:omega_prior}) and the bottom-right plot shows the estimated off-diagonal estimates of the precision matrix when increasing the graphical LASSO penalty parameter. In all plots, \texttt{-o-} lines correspond to true associations in the simulated data, and \texttt{-x-} lines correspond to coefficients representing no underlying association.}
\label{fig:RockovaPlot}
\end{figure}

An advantage of using spike-and-slab priors for covariate and network edge selection, over penalized methods, is given by the flexible level of sparsity induced on the regression coefficients and the precision matrix entries. For example, \cite{Li2017} show that $d^*_{i,j}$ from equation \eqref{eq:d_star update} is comparable to the penalty parameter, $\lambda$, in the graphical LASSO \citep{Dempster}.  However, $d^*_{i,j}$ is unique to each edge and is adaptively learned from the data. In Figure \ref{fig:RockovaPlot} we show this advantage over penalized methods by plotting the estimated coefficients and precision matrix values, for a smaller simulation scenario with $p = 10$, $q = 15$, and $n = 5000$, using SINC (with $\tau$ fixed at 1) and the penalization based method mLDM, while varying the sparsity inducing parameters. The top-left plot shows the estimated $\mathbf{B}$ coefficients by the proposed model when increasing the variance parameter $\nu_B$ of the spike-and-slab prior in equation \eqref{eq:B prior}. The top-right plot shows the estimated $\mathbf{B}$ coefficients via mLDM when increasing the LASSO penalty parameter. The bottom-left plot shows the estimated off-diagonal values of the precision matrix $\mathbf{\Omega}$ when increasing the variance parameter $\nu_0$ in equation \eqref{eq:omega_prior} and the bottom-right plot shows the estimated off-diagonal estimates of the precision matrix when increasing the graphical LASSO penalty parameter. In all plots, red lines correspond to true associations in the simulated data, and black lines correspond to coefficients representing no underlying association. The flat trend in the red lines of the plots related to SINC shows that the estimated covariate effects and precision matrix entries corresponding to true associations are stable, while for mLDM, depicted at bottom, they get shrunken to zero as the penalty parameters increase.

We conclude this section by providing some comments on the sensitivity of the results to the choice of the hyperparameters. As shown in Figure \ref{fig:RockovaPlot}, with sufficient data, the estimates of $\mathbf{\Omega}$ and $\mathbf{B}$ are stable for increasing values of $\nu_0$, in the prior of equation \eqref{eq:omega_prior}, and $\nu_B$, in the prior of equation  \eqref{eq:B prior}, respectively. These parameters, however, affect the sparsity of the selection. In particular, as $\nu_0$ increases, holding all other parameters constant, the selected network becomes sparser. Similarly, as $\nu_1$, which appears in the prior of equation \eqref{eq:omega_prior}, increases, holding $\nu_0$ constant, the network sparsity increases. In recent work using this type of mixture prior, \cite{Li2017} and \cite{RockovaGeorge} have suggested holding $\nu_1$ constant while varying $\nu_0$. It should be apparent, then, that increasing $\nu_B$ while holding all other parameters constant decreases the number of selected coefficients, as increasing $\nu_B$ is analogous to increasing $\nu_1$. The remainder of the hyperparameters influence sparsity as well, but to a lesser extent. For example, changing $a_\pi$ and $b_\pi$ in the prior given in equation \eqref{eq:edge prior} to put more weight on larger values of $\pi$ results in sparser networks. Similarly, selecting $a_\gamma$ and $b_\gamma$ in the prior of equation \eqref{eq:B prior} to reflect a stronger prior belief in larger $\boldsymbol{\theta}_\gamma$ values results in an increase in the number of selected coefficients. Since $\pi$ and $\boldsymbol{\theta}_\gamma$ are both updated at each iteration of the SINC algorithm, selecting relatively non-informative priors, such as the ones used in the simulations of $a_\gamma = b_\gamma = 2$, allows the sparsity levels to be primarily controlled by $\nu_0$ and $\nu_B$. Alternative choices that would also be appropriate include $a_\gamma = b_\gamma = 1$, a more non-informative setting corresponding to a uniform prior on the unit interval, or $a_\gamma = 1$ and $b_\gamma = p$, which would more strongly favor sparsity, as discussed in \cite{RockovaGeorge}. We found that our variable selection results were not overly sensitive to the choice of these parameters. Increasing $\lambda$ also increases the network sparsity because it changes the scale of the estimated $\mathbf{\Omega}$ values by making them smaller. Appropriate $\lambda$ values need to be selected based on the scale of the data that is being used.

\section{Application to a study of the vaginal microbiome in pregnancy}
\label{section:app}

In this section, we apply our proposed method to data from the Multi\textsf{'}Omic Microbiome Study - Pregnancy Initiative (MOMS-PI), an NIH-funded study aimed at characterizing the microbiome and its role in shaping maternal and infant health. Previous research has demonstrated that immune and metabolic changes during pregnancy reshape the microbiome, which undergoes large shifts during the course of pregnancy. The vaginal microbiome in particular has been shown to  change early in pregnancy \citep{Serrano2019} and be predictive of pregnancy outcomes such as  preterm birth \citep{Fettweis2019}. 

The MOMS-PI study involved following pregnant women throughout pregnancy and for a short term after childbirth. Participants in the MOMS-PI study were asked to provide samples from the mouth, skin, vagina and rectum.  Multiple omic technologies were used to process the collected samples including microbiome profiling, metabolomics, and quantification of cytokine abundances via immunoproteomics. Cytokines, in particular, are one mechanism by which the host regulates the composition of the vaginal microbiome. The data was obtained from the R package \texttt{HMP2Data} and consists of 596 subjects that were sampled across multiple visits. For our analysis, we focus on samples collected at the first baseline visit. Of the 596 subjects, 225 subjects had both the microbiome and cytokine profiling of the vagina available at this time point. We consider these 225 subjects in the analysis. To avoid the inclusion of very rare taxa, the OTUs were filtered for inclusion in the analysis using the following rule: the absolute abundance of an OTU had to be greater than 1 for at least 10 percent of the subjects, resulting in 90 OTUs. All 29 cytokines profiled were included as covariates. For the analysis, the cytokine data was transformed to the log scale, and centered.

We applied the SINC method to estimate the interaction between vaginal microbes, as well as the interplay between vaginal cytokines and microbial abundances. We used the same hyperparameter settings as in the simulation: $\nu_B = 1$, $a_\gamma = 2$, $b_\gamma = 2$, $\nu_1 = 10$, $\lambda = 150$, $a_\pi = 2$, $b_\pi = 2$, set $\nu_0 = 0.01$, a value that, in the simulations, achieved a sparsity level of 0.10, and fixed $\tau$ to 1. Since we are controlling for cytokine counts when estimating the microbiome network, we are more confident in the selection as we do not expect to select an edge between two microbes that may be related only via their common dependence on a cytokine.

\subsection{MOMS-PI Results}
\begin{figure}
    \centering
    \textbf{a)}
    \includegraphics[width=3.0in,height=3.0in,page=1] {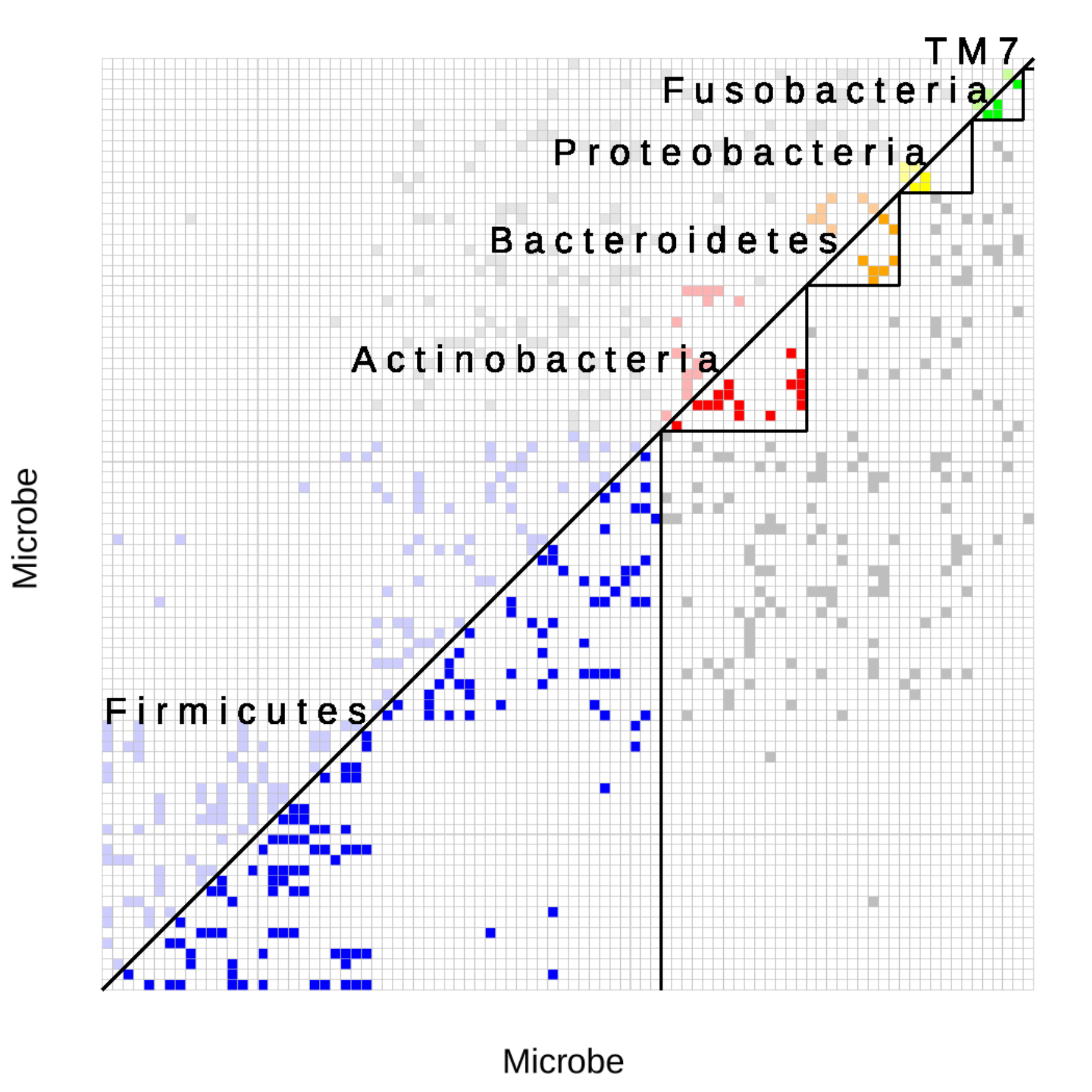}
    \textbf{b)}
    \includegraphics[width=3.0in,height=3.0in,page = 1] {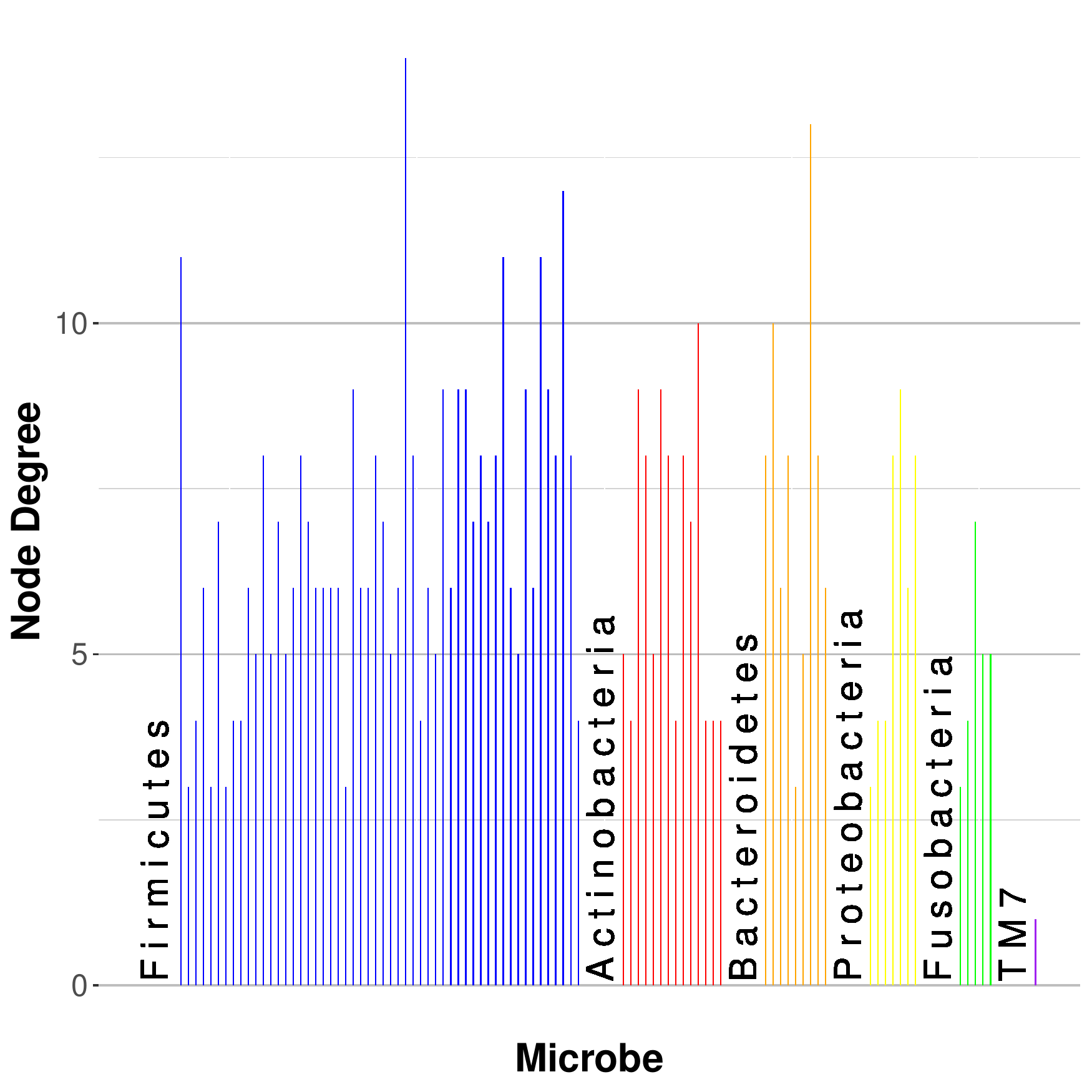}
    \textbf{c)} 
    \includegraphics[width=4.75in,height=4.75in,page = 1]{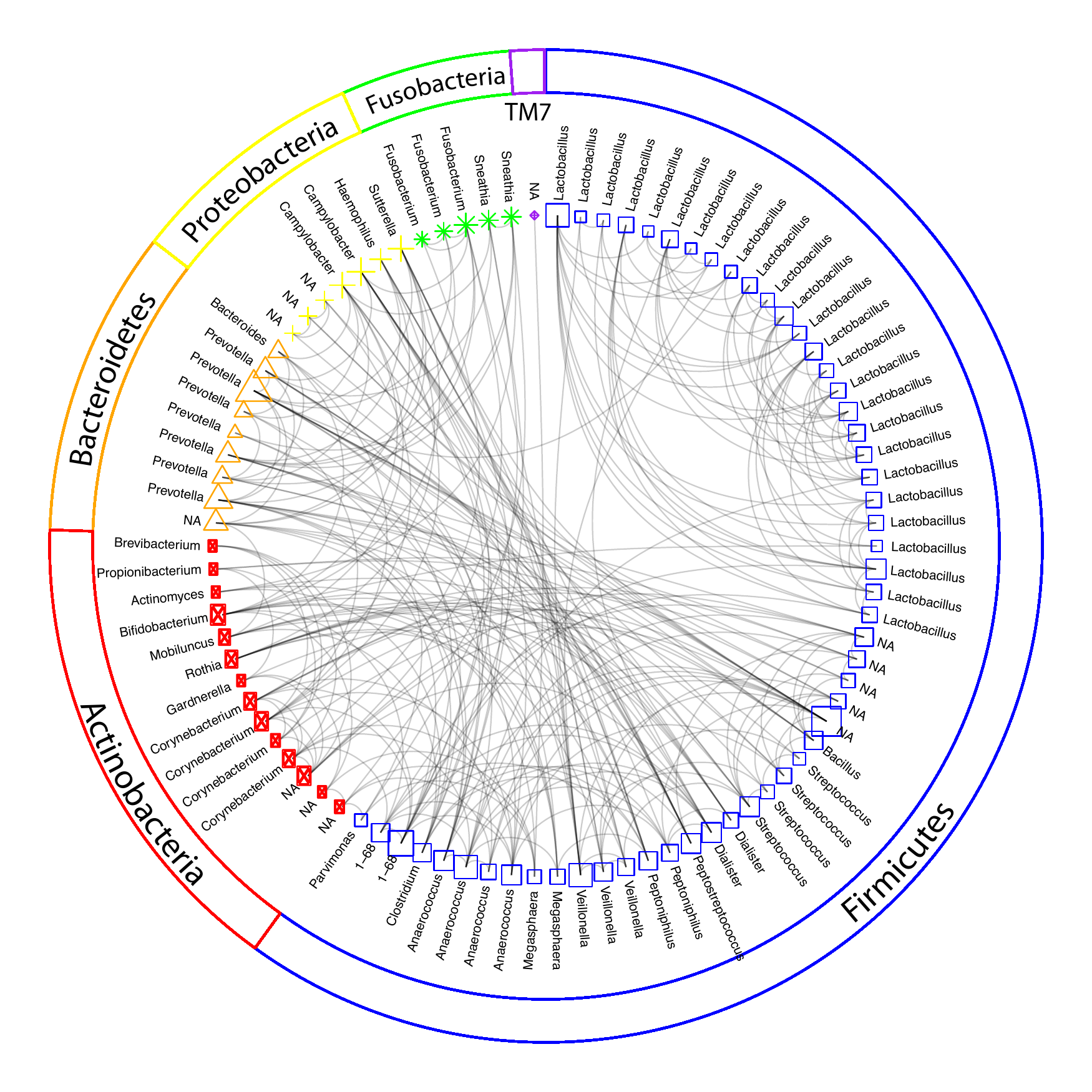}
       \caption{Case study: Plot {\it a)}: Adjacency matrix of the microbial network inferred from the MOMS-PI data, with filled boxes representing selected edges. Plot {\it b)}: number of edges for each OTU, listed in the same order as in the adjacency matrix. Plot {\it c)}: Network diagram of the inferred microbial network, with node sizes representing the degree of the nodes. 
}
\label{fig:Heatmap}
\end{figure}

Figure \ref{fig:Heatmap}a shows the adjacency matrix of the microbial network inferred from the MOMS-PI data, with filled boxes representing selected edges, together with a plot of the number of edges for each OTU in \ref{fig:Heatmap}b. A network diagram of the inferred microbial network is shown in in Figure \ref{fig:Heatmap}c, with node sizes representing the degree of the nodes, so the larger a node, the more edges that node has with other nodes. In these plots, OTUs are grouped based on their phylum (Firmicutes, Actinobacteria, Bacteroidetes, Proteobacteria, Fusobacteria, and TM7). Looking at the adjacency matrix and network representation, we notice that Actinobacteria have few shared edges with Bacteroidetes, Proteobacteria and Fusobacteria, instead sharing the majority of their inter-phylum edges with Firmicutes, while the other phyla (Firmicutes, Bacteroidetes, Proteobacteria and Fusobacteria) show no trend in inter-phylum edges. We also notice that within the Firmicutes subnetwork, OTUs of the genus \textit{Lactobacillus} (OTU 1 through 26 in the adjacency matrix) form their own subnetwork with very few inter-genus connections. Also, from the node degrees plot we can see that many Firmicutes have large numbers of edges. Indeed, when looking at the most connected nodes of the inferred microbial network, we found that 5 of the 6 most connected OTUs belong to the Firmicutes phylum.

\begin{figure}
    \centering
    \includegraphics[width=5.5in,height=3.0in, page = 1]{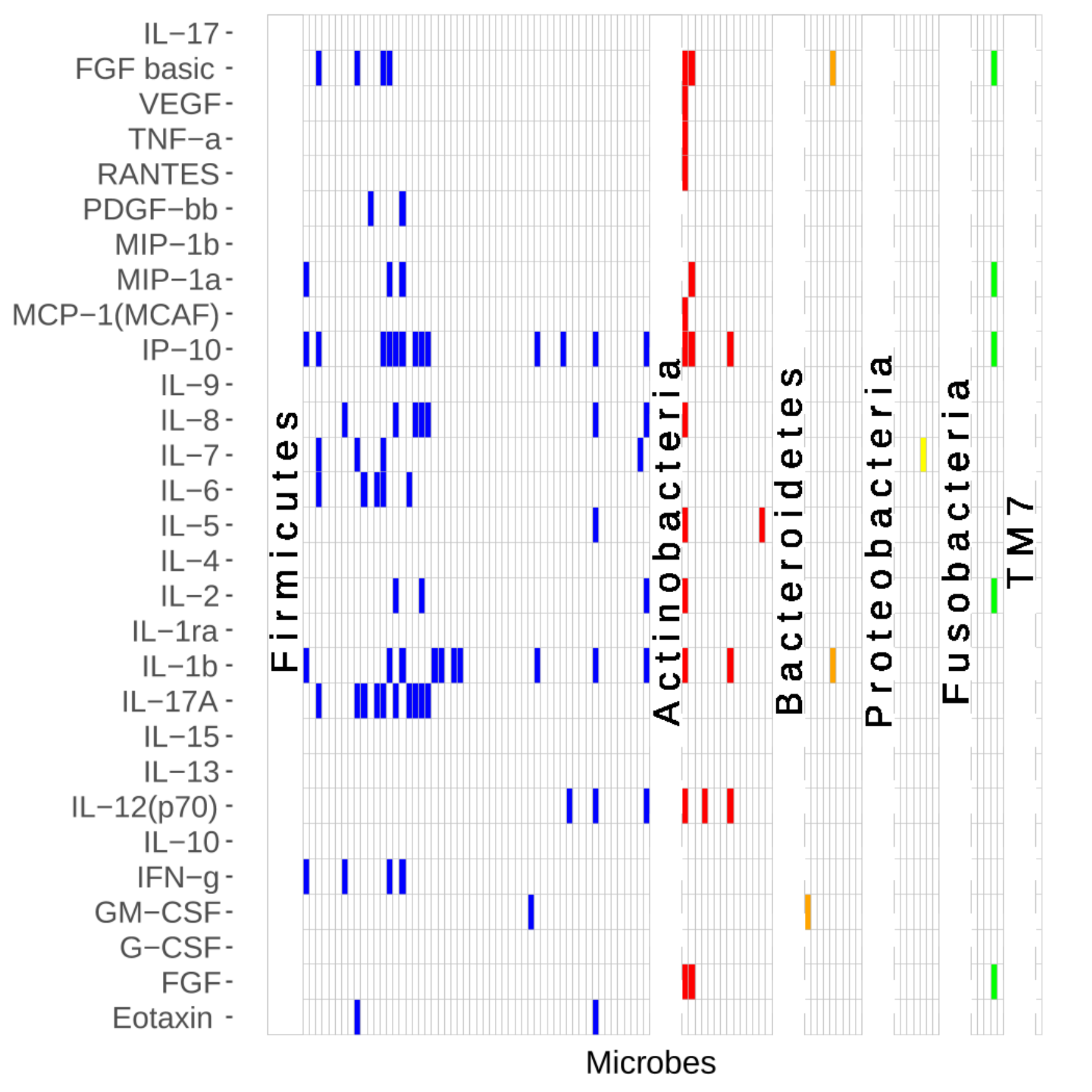}
    \caption{Case study: Adjacency matrix of the cytokine and microbial associations inferred from the MOMS-PI data. Filled boxes indicate that there is an association between a cytokine (plotted on the y axis) and an OTU (plotted on the x axis). 
 }
\label{fig:MicroCyto}
\end{figure}

Next, we show the inference on the microbe-cytokine association network. Figure \ref{fig:MicroCyto} shows the adjacency matrix of the selected microbe-cytokine associations, with microbes colored based on phylum. We observe clear patterns of association, with both cytokines that show relationships to many OTUs, and OTUs that show relationships with several cytokines. In particular, we see that the cytokines IP-10, IL-1b, IL-17A, FGF basic, and IL-8 have the most associations with OTU abundances. When looking at which OTUs have the most associations with cytokines, we found that 6 of the 10 most connected are \textit{Lactobacillus}. This is also seen in Figure \ref{fig:MicroCyto}, where many of the first 26 microbes (columns) have several cytokine associations. \textit{Lactobacillus} has previously been shown to be largely influenced by cytokines \citep{Valenti2018}. 

\begin{figure}
    \centering
    \textbf{a)}
    \includegraphics[height=2.0in, page = 1]{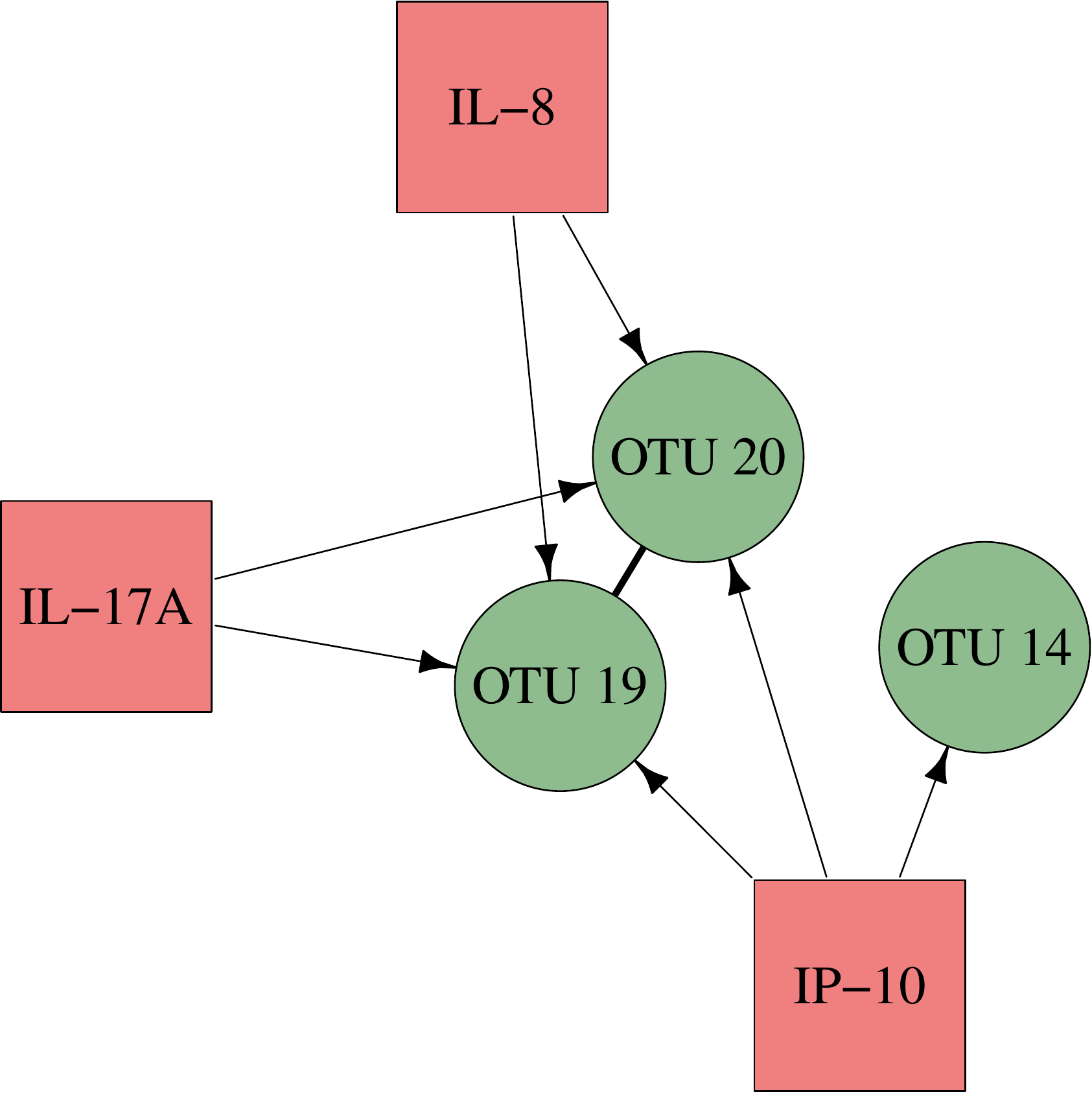}
    \hspace{.5in}
    \textbf{b)}
    \includegraphics[height=1.8in, page = 1]{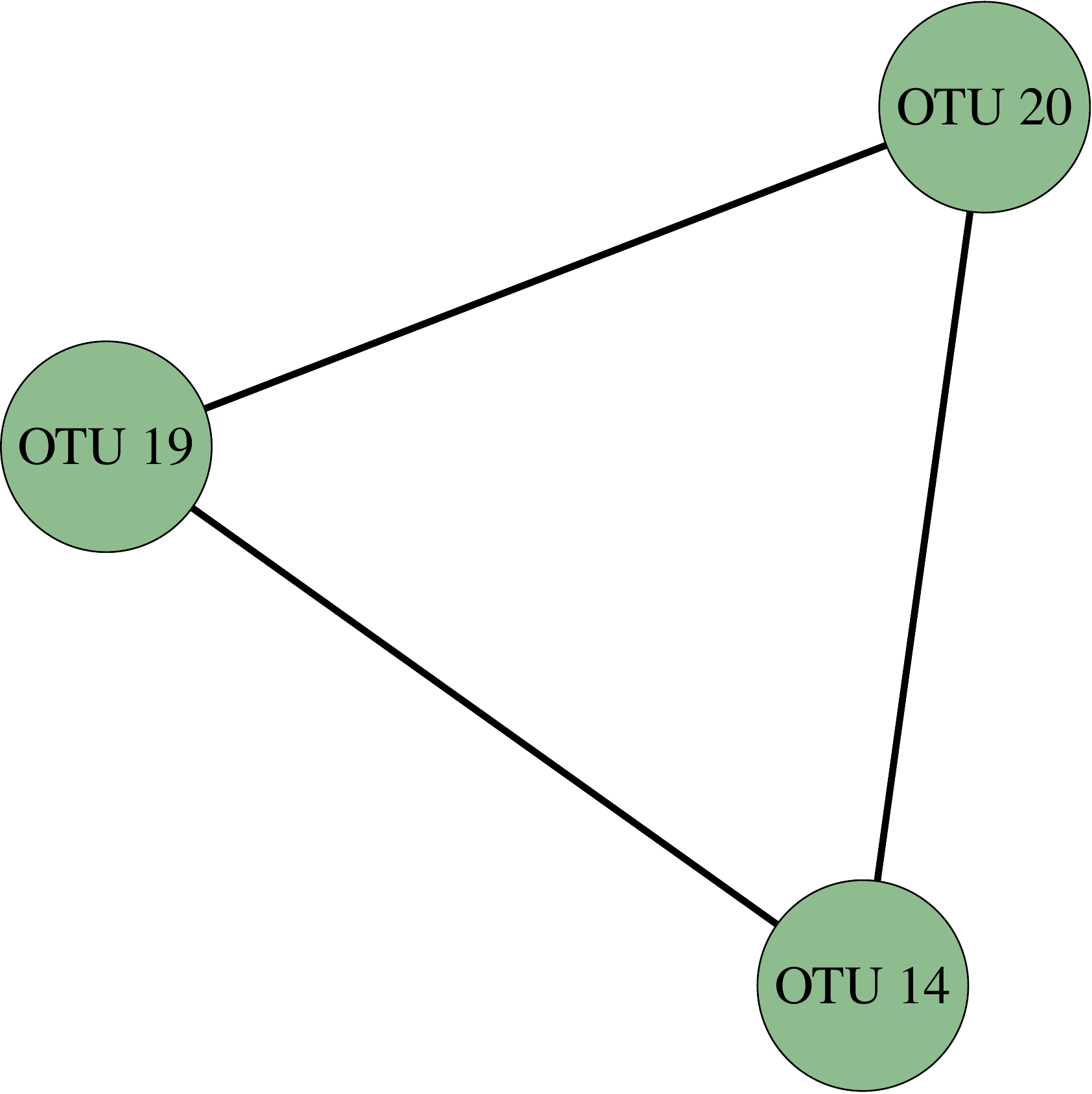}
    \caption{Case study: Plot {\it{a)}} Subnetwork of selected edges in a microbe-microbe network and selected associations between microbes and cytokines, found when using SINC. Plot {\it{b)}} Subnetwork of selected edges in a microbe-microbe network, found when using both SpeicEasi and a variation of SINC with the \textbf{B} fixed to 0. Cytokines are not included in the subnetwork of plot {\it{b)}}, as neither model accounts for cytokines.}
\label{fig:MicroCyto_network}
\end{figure}

We also compare the results from SINC to those from SpiecEasi \citep{spiecEasi} and the \textbf{B}-constrained version of SINC, and found that the two methods that do not control for covariates shared 22 edges that were not selected by the proposed model. Two of these edges can be seen in Figure \ref{fig:MicroCyto_network}, which shows the network of a subset of three microbes and three cytokines estimated by SINC, as well as the network of the same subset of microbes estimated by SpiecEasi and a variant of SINC with the \textbf{B} coefficients not estimated. We hypothesize that SINC did not select the same edges as the other two methods, i.e. the edge between OTUs 14 and 19 and the edge between OTUs 19 and 20, because edges selected by methods that do not control for cytokines may incorrectly determine an edge when microbe pairs have a mutual association with a cytokine. This can be seen in Figure \ref{fig:MicroCyto_network}, where OTUs 14,19, and 20 all have an association with IP-10. This illustrates the ability of our model to discover covariate effects and a sparse network accounting for these effects.

\section{Discussion}
\label{section:disc}

In this paper, we have introduced a novel Bayesian hierarchical model for count data that allows for simultaneous estimation of covariate dependence and network interactions. By accounting for covariate selection, simultaneous estimation methods are able to control for those variables, which ultimately leads to more accurate network estimation. We have considered multivariate count data, and specifically compositional data that have a fixed sum constraint, and have modeled the data using a Dirichlet-Multinomial likelihood. We have accounted for covariates by modeling the log concentration parameters via a Gaussian distribution, and achieved simultaneous covariate and edge selection via spike-and-slab priors.  For posterior inference, we have implemented a variational Bayes approach that includes an expectation-minimization step to enable efficient estimation. We have shown through simulations that the proposed model outperforms existing methods in its accuracy of network recovery. This is due, in part, to the flexibility of the hierarchical model, as discussed in Section \ref{ssec:influence}, which avoids some of the over-shrinkage typical of penalized approaches, as well as the added accuracy from doing simultaneous covariate and network selection. Finally, we have applied the proposed method to data from the Multi-\textsf{'}Omic Microbiome Study-Pregnancy Initiative (MOMS-PI) study, to estimate the microbial interactions in the vagina, as well as the interplay between vaginal cytokines and microbial abundances, providing insight into mechanisms of host-microbial interaction during pregnancy. 

Although other estimation methods, such as EM, could potentially be applied, we found that our unique hybrid algorithm offers several advantages. As we noted in Section 3, the VI and EM approaches are similar in many ways. 
VI, however, can enable additional insight on the uncertainty of the parameter estimates, as one can think of the EM as a special case of the VI algorithm when the variational distributions are point estimates. Although this is one potential advantage of the VI estimation of $\mathbf{B}$, we primarily prefer the VI approach for pragmatic reasons regarding performance, since we found in practice that the VI algorithm for variable selection is less sensitive to the choice of the hyperparameters (specifically, the standard deviations of the spike and slab distributions) than alternative approaches for variable selection in the EM framework. This makes the application of our approach simpler, since we can focus on tuning the parameters for the EMGS portion of the algorithm. Moreover, \cite{ray2019} recently demonstrated that the VI algorithm of \cite{carbonetto} generally outperforms the EMVS algorithm of \cite{RockovaGeorge} across various simulation settings, suggesting we may be able to obtain more accurate estimates of $\mathbf{B}$ under this approach. Finally, VI methods can be used with discrete spike and slab priors, whereas  continuous spike and slab priors are used with EM methods.

Although our VI scheme is more computationally efficient than MCMC sampling, estimating the latent variable matrix $\mathbf{Z}$ is still computationally expensive and a bottleneck to this problem. For the case study of this paper, the model ran on a cluster using 25 cores and took 16 minutes (approximately 6.2 CPU hours). Using the case study data and the default R package settings, mSSL-DPE took 49.4 minutes, SpiecEasi took 57 seconds. Even though spiecEasi and mSSL-DPE were much faster, they resulted in less accurate predictions. Finally, mLDM was much slower and took over 120 hours per simulation. The dramatic difference  in time between mLDM and SINC is due, in part, to SINC being calibrated for parallel computing. Not only does the computational complexity of our model scale in $p$, but because there are $p$ $\times$ $n$ latent variables in $\mathbf{Z}$, the speed of our algorithm also scales with $n$. Avoiding estimation of these latent variables, or finding computationally more efficient estimates, would allow for further scalability of the implementation.

Python code implementing the SINC method is available at \href{https://github.com/Nathan-Osborne/SINC/}{https://github.com/Nathan-Osborne/SINC/}.

\section*{Acknowledgements}
Research supported by NSF/DMS 1811568/1811445.

\bibliographystyle{apalike}
\bibliography{ref}

\begin{thebibliography}{}

\bibitem[Barbieri and Berger, 2004]{Barbieri2004}
Barbieri, M.~M. and Berger, J.~O. (2004).
\newblock Optimal predictive model selection.
\newblock {\em The Annals of Statistics}, 32(3):870--897.

\bibitem[Blei et~al., 2017]{blei2017}
Blei, D.~M., Kucukelbir, A., and McAuliffe, J.~D. (2017).
\newblock Variational inference: A review for statisticians.
\newblock {\em Journal of the American statistical Association},
  112(518):859--877.

\bibitem[Brown et~al., 1998]{brown98}
Brown, P.~J., Vannucci, M., and Fearn, T. (1998).
\newblock Multivariate {B}ayesian variable selection and prediction.
\newblock {\em Journal of the Royal Statistical Society: Series B (Statistical
  Methodology)}, 60(3):627--641.

\bibitem[Carbonetto and Stephens, 2012]{carbonetto}
Carbonetto, P. and Stephens, M. (2012).
\newblock Scalable variational inference for {B}ayesian variable selection in
  regression, and its accuracy in genetic association studies.
\newblock {\em {B}ayesian Analysis}, 7(1):73–108.

\bibitem[Chen and Li, 2013]{chen_li_2013}
Chen, J. and Li, H. (2013).
\newblock Variable selection for sparse {D}irichlet-multinomial regression with
  an application to microbiome data analysis.
\newblock {\em The annals of applied statistics}, 7(1):418--42.

\bibitem[David et~al., 2014]{David2014}
David, L.~A., Maurice, C.~F., Carmody, R.~N., Gootenberg, D.~B., Button, J.~E.,
  Wolfe, B.~E., Ling, A.~V., Devlin, A.~S., Varma, Y., Fischbach, M.~A., et~al.
  (2014).
\newblock Diet rapidly and reproducibly alters the human gut microbiome.
\newblock {\em Nature}, 505(7484):559--563.

\bibitem[Dempster, 1972]{Dempster}
Dempster, A. (1972).
\newblock Covariance selection.
\newblock {\em Biometrics}, 28:157--175.

\bibitem[Deshpande et~al., 2019]{Deshpande2019}
Deshpande, S.~K., Ročková, V., and George, E.~I. (2019).
\newblock Simultaneous variable and covariance selection with the multivariate
  spike-and-slab lasso.
\newblock {\em Journal of Computational and Graphical Statistics}, 0(0):1--11.

\bibitem[Edwards et~al., 2019]{Edwards2019}
Edwards, V.~L., Smith, S.~B., McComb, E.~J., Tamarelle, J., Ma, B., Humphrys,
  M.~S., Gajer, P., Gwilliam, K., Schaefer, A.~M., Lai, S.~K., et~al. (2019).
\newblock The cervicovaginal microbiota-host interaction modulates
  \textsl{Chlamydia trachomatis} infection.
\newblock {\em MBio}, 10(4):e01548--19.

\bibitem[Fan and Li, 2001]{SCAD}
Fan, J. and Li, R. (2001).
\newblock Variable selection via nonconcave penalized likelihood and its oracle
  properties.
\newblock {\em Journal of the American Statistical Association},
  96(456):1348--1360.

\bibitem[Fang et~al., 2017]{gCoda}
Fang, H., Huang, C., Zhao, H., and Deng, M. (2017).
\newblock g{C}oda: Conditional dependence network inference for compositional
  data.
\newblock {\em Journal of Computational Biology}, 24(7):699–708.

\bibitem[Fettweis et~al., 2019]{Fettweis2019}
Fettweis, J.~M., Serrano, M.~G., Brooks, J.~P., Edwards, D.~J., Girerd, P.~H.,
  Parikh, H.~I., et~al. (2019).
\newblock The vaginal microbiome and preterm birth.
\newblock {\em Nature Medicine}, 25(6):1012--1021.

\bibitem[Friedman et~al., 2008]{Friedman2008}
Friedman, J., Hastie, T., and Tibshirani, R. (2008).
\newblock Sparse inverse covariance estimation with the graphical lasso.
\newblock {\em Biostatistics}, 9(3):432--441.

\bibitem[George and McCulloch, 1997]{GeorgeMcCulluch}
George, E.~I. and McCulloch, R.~E. (1997).
\newblock Approaches for {B}ayesian variable selection.
\newblock {\em Statistica Sinica}, 7(2):339--373.

\bibitem[Girvan and Newman, 2002]{Girvan2002}
Girvan, M. and Newman, M.~E. (2002).
\newblock Community structure in social and biological networks.
\newblock {\em Proceedings of the national academy of sciences},
  99(12):7821--7826.

\bibitem[Gloor et~al., 2017]{Gloor}
Gloor, G.~B., Macklaim, J.~M., Pawlowsky-Glahn, V., and Egozcue, J.~J. (2017).
\newblock Microbiome datasets are compositional: And this is not optional.
\newblock {\em Frontiers in Microbiology}, 8:22--24.

\bibitem[Huang et~al., 2016]{Huang2016}
Huang, X., Wang, J., and Liang, F. (2016).
\newblock A variational algorithm for {B}ayesian variable selection.

\bibitem[Jiang et~al., 2019]{R_huge}
Jiang, H., Fei, X., Liu, H., Roeder, K., Lafferty, J., Wasserman, L., Li, X.,
  and Zhao, T. (2019).
\newblock {\em huge: High-Dimensional Undirected Graph Estimation}.
\newblock R package version 1.3.3.

\bibitem[Kook et~al., 2020]{Kook2020}
Kook, J.~H., Vaughn, K.~A., DeMaster, D.~M., Ewing-Cobbs, L., and Vannucci, M.
  (2020).
\newblock {BVAR}-{C}onnect: A variational {B}ayes approach to multi-subject
  vector autoregressive models for inference on brain connectivity networks.
\newblock {\em arXiv preprint arXiv:2006.04608}.

\bibitem[Koslovsky and Vannucci, 2020]{MicroBVS}
Koslovsky, M. and Vannucci, M. (2020).
\newblock Micro{BVS}: {D}irichlet-tree multinomial regression models with
  {B}ayesian variable selection – an {R} package.
\newblock {\em BMC Bioinformatics}, 21:301.

\bibitem[Kurtz et~al., 2015]{spiecEasi}
Kurtz, Z.~D., Müller, C.~L., Miraldi, E.~R., Littman, D.~R., Blaser, M.~J.,
  and Bonneau, R.~A. (2015).
\newblock Sparse and compositionally robust inference of microbial ecological
  networks.
\newblock {\em PLOS Computational Biology}, 11:1--25.

\bibitem[Lenkoski and Dobra, 2011]{Lenkoski2011}
Lenkoski, A. and Dobra, A. (2011).
\newblock Computational aspects related to inference in {G}aussian graphical
  models with the {$G$}-{W}ishart prior.
\newblock {\em J. Comput. Graph. Stat.}, 20(1):140--157.

\bibitem[Li et~al., 2020]{LiMcCormick20}
Li, Z.~R., McComick, T.~H., and Clark, S.~J. (2020).
\newblock Using {B}ayesian latent {G}aussian graphical models to infer symptom
  associations in verbal autopsies.
\newblock {\em Bayesian Analysis}, to appear.

\bibitem[Li and McCormick, 2019]{Li2017}
Li, Z.~R. and McCormick, T.~H. (2019).
\newblock An expectation conditional maximization approach for {G}aussian
  graphical models.
\newblock {\em Journal of Computational and Graphical Statistics}, To appear.

\bibitem[Meinshausen and B\"uhlmann, 2006]{Meinshausen}
Meinshausen, N. and B\"uhlmann, P. (2006).
\newblock High-dimensional graphs and variable selection with the lasso.
\newblock {\em Ann. Statist.}, 34(3):1436--1462.

\bibitem[Miao et~al., 2019]{miao}
Miao, Y., Kook, J., Lu, Y., Guindani, M., and Vannucci, M. (2019).
\newblock Scalable {B}ayesian variable selection regression models for count
  data.
\newblock In F., Y., M., S., D., N., and J.-L., D.-B., editors, {\em Flexible
  {B}ayesian Regression Modelling}, pages 187--219. Elsevier.

\bibitem[Miao et~al., 2020]{miao2020}
Miao, Y., Kook, J.~H., Lu, Y., Guindani, M., and Vannucci, M. (2020).
\newblock Scalable bayesian variable selection regression models for count
  data.
\newblock In {\em Flexible Bayesian Regression Modelling}, pages 187--219.
  Elsevier.

\bibitem[Mitchell and Beauchamp, 1988]{mitchell_beauchamp_1988}
Mitchell, T.~J. and Beauchamp, J.~J. (1988).
\newblock {B}ayesian variable selection in linear regression.
\newblock {\em Journal of the American Statistical Association},
  83(404):1023--1032.

\bibitem[Ntzoufras et~al., 2003]{Ntzoufras2003}
Ntzoufras, I., Dellaportas, P., and Forster, J.~J. (2003).
\newblock {B}ayesian variable and link determination for generalised linear
  models.
\newblock {\em Journal of Statistical Planning and Inference},
  111(1-2):165--180.

\bibitem[Raftery, 1996]{Raftery1996}
Raftery, A.~E. (1996).
\newblock Approximate {B}ayes factors and accounting for model uncertainty in
  generalised linear models.
\newblock {\em Biometrika}, 83(2):251--266.

\bibitem[Ray and Szabo, 2019]{ray2019}
Ray, K. and Szabo, B. (2019).
\newblock Variational bayes for high-dimensional linear regression with sparse
  priors.
\newblock {\em arXiv preprint arXiv:1904.07150}.

\bibitem[Richardson et~al., 2010]{richardson2010}
Richardson, S., Bottolo, L., and Rosenthal (2010).
\newblock {B}ayesian models for sparse regression analysis of high dimensional
  data.
\newblock In {\em {B}ayesian Statistics 9}, pages 539--569.

\bibitem[Rothman et~al., 2010]{Rothman2010}
Rothman, A.~J., Levina, E., and Zhu, J. (2010).
\newblock Sparse multivariate regression with covariance estimation.
\newblock {\em Journal of Computational and Graphical Statistics},
  19(4):947--962.
\newblock PMID: 24963268.

\bibitem[Roverato, 2002]{Roverato}
Roverato, A. (2002).
\newblock Hyper inverse {W}ishart distribution for non-decomposable graphs and
  its application to {B}ayesian inference for {G}aussian graphical models.
\newblock {\em Scandinavian Journal of Statistics}, 29(3):391--411.

\bibitem[Ročková and George, 2014]{RockovaGeorge}
Ročková, V. and George, E.~I. (2014).
\newblock Emvs: The {EM} approach to {B}ayesian variable selection.
\newblock {\em Journal of the American Statistical Association},
  109(506):828--846.

\bibitem[Serrano et~al., 2019]{Serrano2019}
Serrano, M.~G., Parikh, H.~I., Brooks, J.~P., Edwards, D.~J., Arodz, T.~J.,
  Edupuganti, L., Huang, B., Girerd, P.~H., Bokhari, Y.~A., Bradley, S.~P.,
  et~al. (2019).
\newblock Racioethnic diversity in the dynamics of the vaginal microbiome
  during pregnancy.
\newblock {\em Nature Medicine}, page~1.

\bibitem[Sha et~al., 2004]{Sha2004}
Sha, N., Vannucci, M., Tadesse, M.~G., Brown, P.~J., Dragoni, I., Davies, N.,
  et~al. (2004).
\newblock {B}ayesian variable selection in multinomial probit models to
  identify molecular signatures of disease stage.
\newblock {\em Biometrics}, 60(3):812--819.

\bibitem[Stingo et~al., 2010]{stingo2010}
Stingo, F., Chen, Y., Vannucci, M., Barrier, M., and Mirkes, P. (2010).
\newblock A {B}ayesian graphical modeling approach to micro{RNA} regulatory
  network inference.
\newblock {\em Ann. Appl. Stat.}, 4(4):2024--2048.

\bibitem[Talhouk et~al., 2012]{Talhouk12}
Talhouk, A., Doucet, A., and Murphy, K. (2012).
\newblock Efficient {B}ayesian inference for multivariate probit models with
  sparse inverse correlation matrices.
\newblock {\em Journal of Computational and Graphical Statistics},
  21(3):739--757.

\bibitem[Tang et~al., 2018]{LiMa}
Tang, Y., Ma, L., and Nicolae, D. (2018).
\newblock {A phylogenetic scan test on {D}irichlet-tree multinomial model for
  microbiome data}.
\newblock {\em Annals of Applied Statistics}, 12(1):1--26.

\bibitem[Tibshirani, 1996]{LASSO}
Tibshirani, R. (1996).
\newblock Regression shrinkage and selection via the lasso.
\newblock {\em Journal of the Royal Statistical Society. Series B
  (Methodological)}, 58(1):267--288.

\bibitem[Titsias and L{\'a}zaro-Gredilla, 2011]{Titsias2011}
Titsias, M.~K. and L{\'a}zaro-Gredilla, M. (2011).
\newblock Spike and slab variational inference for multi-task and multiple
  kernel learning.
\newblock In {\em Advances in Neural Information Processing Systems}, pages
  2339--2347.

\bibitem[Valenti et~al., 2018]{Valenti2018}
Valenti, P., Rosa, L., Capobianco, D., Lepanto, M.~S., Schiavi, E., Cutone, A.,
  Paesano, R., and Mastromarino, P. (2018).
\newblock Role of lactobacilli and lactoferrin in the mucosal cervicovaginal
  defense.
\newblock {\em Frontiers in Immunology}, 9:376.

\bibitem[Vinci et~al., 2018]{Vinci18}
Vinci, G., Ventura, V., Smith, M., and Kass, R. (2018).
\newblock Adjusted regularization in latent graphical models: Application to
  multiple-neuron spike count data.
\newblock {\em Annals of Applied Statistics}, 12(2):1068--1095.

\bibitem[Wadsworth et~al., 2017]{Wadsworth}
Wadsworth, W.~D., Argiento, R., Guindani, M., Galloway-Pena, J., Shelburne,
  S.~A., and Vannucci, M. (2017).
\newblock An integrative {B}ayesian {D}irichlet-multinomial regression model
  for the analysis of taxonomic abundances in microbiome data.
\newblock {\em BMC Bioinformatics}, 18(1):94.

\bibitem[Wang, 2012]{WangBGLasso}
Wang, H. (2012).
\newblock {B}ayesian graphical lasso models and efficient posterior
  computation.
\newblock {\em {B}ayesian Analysis}, 7(2):771--790.

\bibitem[Wang, 2015]{Wang2015}
Wang, H. (2015).
\newblock Scaling it up: Stochastic search structure learning in graphical
  models.
\newblock {\em {B}ayesian Analysis}, 10(2):351--377.

\bibitem[Yang et~al., 2017]{mLDM}
Yang, Y., Chen, N., and Chen, T. (2017).
\newblock Inference of environmental factor-microbe and microbe-microbe
  associations from metagenomic data using a hierarchical {B}ayesian
  statistical model.
\newblock {\em Cell Systems}, 4(1):129 -- 137.

\bibitem[Yuan and Lin, 2007]{Yuan2007}
Yuan, M. and Lin, Y. (2007).
\newblock Model selection and estimation in the {G}aussian graphical model.
\newblock {\em Biometrika}, 94(1):19--35.

\bibitem[Zou, 2006]{Adaptive}
Zou, H. (2006).
\newblock The adaptive lasso and its oracle properties.
\newblock {\em Journal of the American Statistical Association},
  101(476):1418--1429.

\end{thebibliography}

\end{document}